# Monitoring the evolution of relative product populations at early times during a photochemical reaction

Joao Pedro Figueira Nunes[†1], Lea Maria Ibele[†2], Shashank Pathak[†3], Andrew R. Attar[4=], Surjendu Bhattacharyya[3], Rebecca Boll[5], Kurtis Borne[3], Martin Centurion[1], Benjamin Erk[6], Ming-Fu Lin[4], Ruaridh J.G. Forbes[4], Nate Goff[7], Christopher S. Hansen[8], Matthias Hoffmann[4], David M.P. Holland[9], Rebecca A. Ingle[10], Duan Luo[4], Sri Bhavya Muvva[1], Alex Reid[4], Arnaud Rouzée[11], Artem Rudenko[3], Sajib Kumar Saha[1], Xiaozhe Shen[4], Anbu Selvam Venkatachalam[3], Xijie Wang[4], Matt R. Ware[4#], Stephen P. Weathersby[4], Kyle Wilkin[1], Thomas J.A. Wolf[4,12], Yanwei Xiong[1], Jie Yang[4$], Michael N. R. Ashfold*[13], Daniel Rolles*[3], Basile F. E. Curchod*[13]

[1]*University of Nebraska–Lincoln, Lincoln, NE, USA;* [2]*CNRS, Institut de Chimie Physique UMR8000, Université Paris-Saclay, Orsay, France;* [3]*J.R. Macdonald Laboratory, Physics Department, Kansas State University, Manhattan, KS, USA;* [4]*SLAC National Accelerator Laboratory, Menlo Park, CA, USA;* [5]*European XFEL, Schenefeld, Germany;* [6]*Deutsches Elektronen Synchrotron DESY, Hamburg, Germany;* [7]*Brown University, Providence, RI, USA;* [8]*School of Chemistry, University of New South Wales, Sydney, NSW, Australia;* [9]*Daresbury Laboratory, Warrington, UK;* [10]*Department of Chemistry, University College London, London UK,* [11]*Max Born Institute, Berlin, Germany;* [12]*Stanford PULSE Institute, SLAC National Accelerator Laboratory, Menlo Park, CA, USA;* [13]*School of Chemistry, University of Bristol, Bristol, UK*

**ABSTRACT:** Identifying multiple rival reaction products and transient species formed during ultrafast photochemical reactions and determining their time-evolving relative populations are key steps towards understanding and predicting photochemical outcomes. Yet, most contemporary ultrafast studies struggle with clearly identifying and quantifying competing molecular structures/species amongst the emerging reaction products. Here, we show that mega-electronvolt ultrafast electron diffraction in combination with *ab initio* molecular dynamics calculations offer a powerful route to determining *time-resolved* populations of the various isomeric products formed after UV (266 nm) excitation of the five-membered heterocyclic molecule 2(5H)-thiophenone. This strategy provides experimental validation of the predicted high (~50%) yield of an episulfide isomer containing a strained 3-membered ring within ~1 ps of photoexcitation and highlights the rapidity of interconversion between the rival highly vibrationally excited photoproducts in their ground electronic state.

## INTRODUCTION

Photochemistry addresses the consequences of molecules interacting with light. The field includes studies of the transformations a molecule can undergo following electronic excitation by absorbing an ultraviolet (UV) or visible photon. The desire to understand the formation mechanisms of products in photochemical processes – the photoproducts – has helped stimulate the development of a plethora of time-resolved spectroscopic and theoretical tools for investigating the dynamics of excited state molecules.[1-8]

Some of these techniques are particularly suitable for probing ultrafast processes that occur in the excited electronic state(s) populated by photoexcitation. Such studies can provide rich information about the dynamics and timescales of photoinduced product formation. This should come as no surprise, given that one of the founding tenets of photochemistry is that products are formed as a result of photoinduced changes in the electronic configuration of a molecule (*i.e.*, products arise from excited electronic states).

More generally, however, photoproducts are formed after the photoexcited parent molecule has returned to its ground electronic state. The internal (vibrational) energy gained by the molecule during non-radiative decay from the photoexcited state to the ground state via nonadiabatic coupling at conical intersections between the respective potential energy surfaces (PESs) will almost certainly be dynamically determined (*i.e.*, non-statistically distributed amongst the normal modes) when the molecule first re-appears in the ground state.[9] The anharmonicity of the ground state potential will encourage intramolecular vibrational redistribution (IVR) and the nascent vibrational energy distribution will evolve towards a more statistical microcanonical distribution over all internal modes. But, in the absence of collisions – as is the case in a low-pressure gas phase sample – the molecule cannot dissipate this energy. Any full understanding of photoinduced reaction dynamics in such cases thus also requires techniques that inform on the timescales of this vibrational energy flow, and how (or whether) this evolving distribution of vibrational energy affects photoproduct formation. Deciphering the formation of photoproducts in the ground electronic state following a nonradiative decay requires a robust strategy to monitor the respective photoproduct populations in *real time*.

Time-resolved photoelectron spectroscopy (TRPES) is one technique that is very well suited to following the early-time dynamics of photoexcited molecular systems.[10-16] TRPES is, however, challenged both by the high photon energies (short wavelengths) required to probe molecules in their ground states and by the potential lack of selectivity when it comes to distinguishing photoproducts with similar electronic structures. Intensive theoretical calculations are often required to interpret experimental TRPES signals and extract information on the photoproducts, as in our recent ultrafast TRPES studies of the UV photoinduced ring-opening of 2(5H)-thiophenone.[17]

The sulfur-containing five-member heterocycle 2(5H)-thiophenone displays a prototypical photochemical response upon UV light absorption[18, 19] (Scheme 1): a fast ring-opening process wherein one C–S bond breaks to form a ring-opened (acyclic) form and triggers an ultrafast decay towards the ground ($S_0$) electronic state, which is accessed within 300 fs.[17] The full photochemistry of 2(5H)-thiophenone involves much more than just this simple primary bond fission process, however. Further intra-molecular rearrangements within the vibrationally excited ground state species could potentially lead to reformation of a thiophenone (the 2(5H)- or 2(3H)-isomers) and/or isomerization to various ketenes. Among these ketenes, an exotic episulfide species, 2-(2-thiiranyl)ketene, involving an S-containing three-member ring, has been predicted to dominate at early times following the nonradiative deexcitation (Scheme 1).[17, 20, 21]

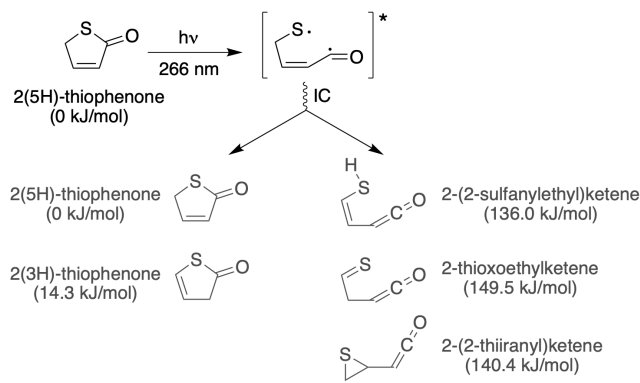

Scheme 1. Photochemistry of 2(5H)-thiophenone upon excitation at 266 nm. Ring-opening takes place in the excited electronic states leading to the formation of a biradical, which undergoes internal conversion (IC) to the ground electronic state where a range of possible photoproducts (dark grey) can be formed with high internal energies. Ring closure can take place, reforming the parent 2(5H)-thiophenone molecule or, potentially, 2(3H)-thiophenone. Alternatively, different ketenes can be formed, including the episulfide (2-(2-thiiranyl)ketene). The relative electronic energies for key structures are given between parenthesis (calculated at the CCSD(T)-F12/cc-pVDZ-F12//MP2/6-311+G\*\* level of theory).

The formation of a three-membered ring from a five-membered cyclic species following the absorption of a 266-nm photon, with 4.65 eV of energy, might appear to challenge chemical intuition. Our recent TRPES studies using suitably short probe wavelengths[17] succeeded in resolving the ultrafast non-radiative decay of 2(5H)-thiophenone from its second excited singlet electronic state ($S_2$, with $n(S)\pi^*$ character in the Franck-Condon region) following photoexcitation at 266 nm (Fig. 1A) and in revealing the formation of acyclic photoproducts. However, an unambiguous identification of the various products was limited by their very similar low-energy ionization potentials – all of which are associated with removing an electron from a relatively unperturbed sulfur lone pair – thus preventing definitive identification of the predicted episulfide.[17, 20]

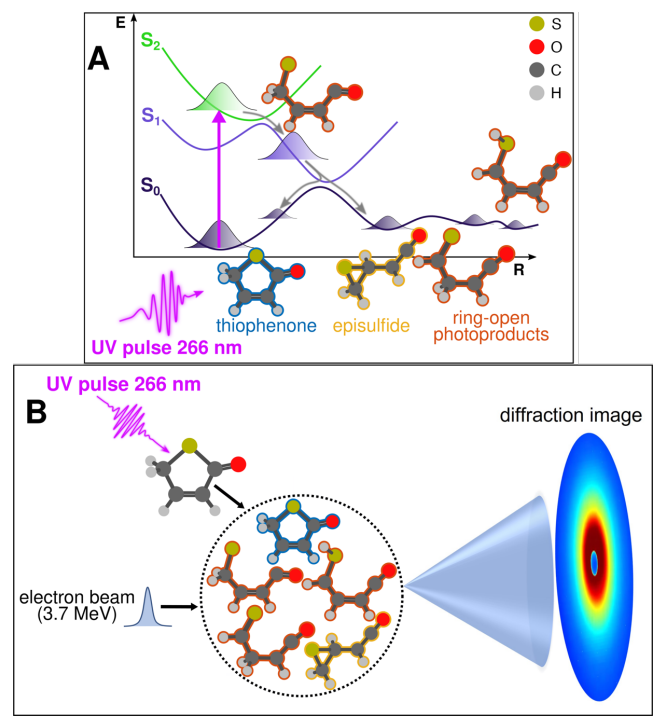

Figure 1. Schematic representation of the photochemistry of 2(5H)-thiophenone and the UED experiment. (**A**): Summary of 2(5H)-thiophenone photochemistry following irradiation at 266 nm. Excitation to the second excited singlet electronic state ($S_2$, green potential energy curve, plotted as a function of a generalized reaction coordinate **R**) results in immediate C–S bond extension (ring opening) and ultrafast nonradiative decay via the $S_1$ (blue) to the ground ($S_0$, dark purple) electronic states. The population of photoexcited 2(5H)-thiophenone molecules fully returns to the $S_0$ state within 300 fs, whereupon the athermal dynamics drives formation of different families of photoproducts (presented in Scheme 1) – all with a substantial internal energy. (**B**): Schematic of the experimental setup. An ultrashort duration UV (266 nm) pulse photoexcites an ensemble of gas-phase 2(5H)-thiophenone molecules inside the UED target chamber, which is interrogated by an incident 3.7 MeV electron beam yielding a diffraction pattern in reciprocal space that is recorded on a position-sensitive detector.

Here, we demonstrate that the sub-picosecond time resolution offered by the mega-electron volt ultrafast electron diffraction (MeV-UED) at the SLAC National Accelerator Laboratory,[3, 22-25] again working hand in hand with theoretical chemistry, offers a very powerful complement – capable of identifying and distinguishing different product families following UV photoinduced ring-opening of 2(5H)-thiophenone and tracking the evolving photoproduct populations. The 2(5H)-thiophenone molecule is first excited with a UV pulse (266 nm) and then probed at different time delays with an electron beam at 3.7 MeV (Fig. 1B). Analyzing the diffraction pattern of the electron beam provides a snapshot of the molecular geometries present at specific time delays following the UV pump pulse. The complementary combination of *ab initio* nonadiabatic and adiabatic

Born-Oppenheimer molecular dynamics simulations allow calculation of photoproduct-family-specific basis functions, which are used to decompose the measured MeV-UED signals and determine the time-dependent relative populations of the various photoproducts.

## METHODS

*Experimental*

The experiment was performed at the MeV UED facility at SLAC National Accelerator Laboratory. The experimental setup (see Fig. S1 in the Supporting Information (SI)) is described in detail elsewhere.[26] Briefly, the output of an 800-nm Ti:Sapphire laser was split into two beams, each of which was frequency-tripled to generate femtosecond ultraviolet (UV) pulses. One of the UV beams was used as a 'pump' to excite the 2(5H)-thiophenone molecules, while the other was used to generate ultrashort electron pulses by irradiating the photocathode of a radio frequency gun. The electron bunches of <150 fs (full width half maximum, FWHM) duration [26] containing ~$10^4$ electrons were accelerated to 3.7 MeV and focused to a spot size of 200 μm FWHM in the interaction region of the gas-phase experimental chamber, where they interacted with the 2(5H)-thiophenone molecules that were delivered into the high-vacuum interaction region in a continuous flow gas cell 3 mm in length with 550 μm openings. 2(5H)-thiophenone (98% purity) was purchased from abcr GmbH and used without further purification. Since 2(5H)-thiophenone has a low vapor pressure (<1 Torr) at room temperature, the sample reservoir and delivery assembly were heated to 60 °C. The pump pulses (15 μJ, 266 nm, ~67 fs (FWHM) in duration) were focused into the interaction chamber to a diameter of 240 μm FWHM and were overlapped with the electron pulses at a 1° crossing angle using a holey mirror. Both the pump laser pulses and the MeV electron pulses were delivered at a repetition rate of 360 Hz. The scattered electrons were detected on a P43 phosphor screen, which was imaged by an Andor iXon Ultra 888 EMCCD camera. Diffraction data were acquired at 37 unique time delays between -3 and 10 ps (Fig. S2). Each time delay was visited a single time per scan and the order in which delays were visited was randomized between scans to minimize systematic errors. In each scan, diffraction data was acquired for 10 seconds (3600 electron shots) at each time delay. A total of 178 scans were acquired, yielding a total integration time per time delay of ~30 minutes. The instrument response function (IRF) of the MeV UED apparatus [26] as used in the present pump-probe configuration is estimated to be 230 fs (FWHM, see Figure S9A in the SI). Additional data validating the assumption that the photochemical findings reported herein are the result of single UV photon excitation processes are presented in the SI (Figs. S11 and S12).

*Nonadiabatic and ab initio molecular dynamics*

The nonadiabatic molecular dynamics (NAMD) following instantaneous photoexcitation of 2(5H)-thiophenone were simulated using Tully's fewest-switches trajectory surface hopping[27] method with the SHARC program package[28, 29] as described in Ref. [17]. 43 initial conditions, sampled from a Wigner distribution for uncoupled harmonic oscillators, were initialized in the bright $S_2$ state. The electronic structure was described at the SA(4)-CASSCF(10/8)/6-31G* level of theory using the Molpro 2012 program package.[30, 31] This level of theory was thoroughly benchmarked in Ref. [17] against XMS-CASPT2. The energy based decoherence correction scheme was employed.[32, 33] After a surface hop, the kinetic energy was rescaled isotropically. A nuclear time step of 0.5 fs was used for all NAMD trajectories. During the excited-state dynamics, the total energy along each trajectory was strictly conserved. Upon deactivation to the ground state, each NAMD trajectory was propagated further until it left the region of strong nonadiabaticity between the ground and first excited electronic state. At this point, the NAMD trajectory was stopped and (ground-state) *ab initio* Born-Oppenheimer molecular dynamics (BOMD) initiated, using the last step of the NAMD trajectory to define the initial conditions. The BOMD trajectories were propagated on the ground electronic state using unrestricted DFT employing the PBE0 exchange/correlation functional[34] and the 6-31G* basis set, with the GPU-accelerated software TeraChem.[35, 36] The BOMD simulations were carried out until a total simulation time (the sum of NAMD and BOMD simulation times) of 2 ps, using a reduced time step of 0.1 fs. A benchmark of this strategy as well as a discussion of the (minimal) influence of triplet states and intersystem crossing is presented in the SI of Ref. [17].

The geometries captured by the (NA+BO)MD simulations were classified into photoproduct categories according to the decision trees shown in Fig. S4 in the SI. This classification relies on the identification of characteristic atomic connectivities using bond lengths or angles. The decision tree for classification I, shown in Fig. S4A, allows the identification of all unique photoproducts captured in the (NA+BO)MD simulations and is inspired by the decision tree reported in Ref. [17]. The scattering signals produced by the photoproducts identified under classification I, depicted in Fig. S5A in the SI, show that ring-opened geometries produce very similar signatures, which cannot be distinguished given the uncertainties within the experimental signal. Therefore, in classification II, shown in Fig. S4B, the ring-opened photoproducts (the excited ring-opened form of 2(5H)-thiophenone and all the acyclic ketene products, see Scheme 1) were grouped under a single classification, *ring-opened*, to ensure that the scattering signatures of all classified photoproducts can be unambiguously distinguished in the experimental signal. The scattering signatures of photoproducts identified under classification II are shown in Fig. S5B (SI).

*Analysis of electron diffraction data and determination of photoproduct branching ratios*

Full details of the electron diffraction data analysis and the routes to extracting photoproduct branching ratios are described in the SI. Briefly, the two-dimensional diffraction patterns recorded at the EMCCD detector were processed into one-dimensional scattering intensity curves, $I(s)$, where $s$ is the momentum transfer vector, which were then decomposed into atomic and molecular scattering contributions. As usual, the diffraction data were then converted to modified scattering intensities, $sM(s)$, to enhance the oscillations in the molecular scattering term and suppress the rapid drop in scattering intensity as a function of $s$ imparted by the $s^{-2}$ scaling in the elastic scattering amplitude. Approximating the $sM(s)$ curve as a sum of sine waves (arising from all internuclear distances in the target molecule) allowed its decomposition into a pair-distribution function (PDF) of all contributing interatomic distances.

Analysis of the time-resolved experimental UED data was based on the difference-diffraction method,[37] wherein the fractional change signal, $\Delta I/I(s,t)$ is defined as

$$\Delta I/I(s,t) = \frac{I(s,t) - I(s,t<0)}{I(s,t<0)}, \quad \text{(eq. 1)}$$

where $I(s,t<0)$ is the reference diffraction signal taken before the arrival of the pump pulse and $I(s,t)$ is the diffraction intensity recorded at pump-probe delay $t$. The experimental time-dependent difference pair distribution functions, $\Delta PDF(r,t)$, shown in Fig. 2 were calculated by applying the sine-transform of the time-dependent difference-modified scattering curves.

The theoretical static $PDF(r)$ and time-dependent $\Delta I/I(s,t)$ and $\Delta PDF(r,t)$ signals reported here were calculated within the independent atom model[38, 39] (IAM) using the nuclear configurations captured along the 43 (NA+BO)MD trajectories. The relative populations for the photoproducts were retrieved directly from the UED signal by fitting a linear combination of basis functions (for thiophenone, acyclic (ring-opened) ketene and episulfide products) selected to reflect the average scattering signatures of these three photoproduct families to the experimental difference-diffraction signal, $\Delta I/I(s,t)$. These photoproduct basis functions were obtained by averaging the theoretical $\Delta I/I(s,t)$ signals for thiophenone, ring-opened and episulfide products in the range $1 \leq t \leq 2$ ps. This approach allows determination of the relative abundances of these three basis functions and their evolution across the experimental time window. These returned relative abundances correspond to relative product populations provided the IAM is valid, i.e., that each isomer of a common ground-state species (neutral $C_4H_4SO$ in this case) shows the same elastic diffraction intensity.

## RESULTS AND DISCUSSION

Steady-state and time-resolved atomic pair distribution functions
The $\Delta PDF(r,t)$ maps obtained from the UED measurements following 266-nm photoexcitation of 2(5H)-thiophenone at pump-probe time delays out to $t = 2$ ps (Fig. 2B) reveal an obvious decrease in the most intense feature in the parent static PDF (Fig. 2A). The theoretical $\Delta PDF(r,t)$ map (Fig. 2C) calculated from a swarm of combined nonadiabatic and ground-state dynamics trajectories (see Methods) is in excellent agreement with the experimental data, reproducing all the main features. The experimental and theoretical $\Delta PDF(r,t)$ maps both show an obvious photoinduced feature ($\delta$, dashed black line in Fig. 2), but the $\alpha$, $\beta$, and $\gamma$ features show seemingly different responses to photoexcitation. The first two, which contain substantial contributions from, respectively, the ring-closed parent C…S (including the C1–S bond that breaks) and S…O separations, both show predictable decreases when some of the parent population is converted to photoproducts, but the $\gamma$ feature shows no similar decrease. As we now show, these differences and the entire $\Delta PDF(r,t)$ maps are completely understandable by recognizing that parent depletion (by photoexcitation) is quickly followed by photoproduct formation once the excited molecules reappear in the $S_0$ state with athermal population and internal energy distributions. The feature labelled $\delta$ in the $\Delta PDF(r,t)$ map already hints at the formation of the predicted episulfide, but additional analysis is required to make this assignment unambiguous and to quantify the episulfide yield.

Monitoring the appearance of the episulfide molecule
To pinpoint the presence of the episulfide product, it proves more revealing to work in terms of $\Delta I/I(s,t)$ and $s$, where $\Delta I$ is the photoinduced change in radially averaged diffraction intensity relative to that from the unexcited sample (i.e., that recorded when the UED probe pulse preceded the 266-nm pump pulse) and $s$ is the momentum transfer vector (see Methods and the SI for additional information). Figs. 3C and 3A show, respectively, the $\Delta I/I(s,t)$ maps measured by UED and calculated from the structures returned by the *ab initio* nonadiabatic and ground state dynamics simulations for the time delay range $-0.5 \leq t \leq +2.0$ ps.

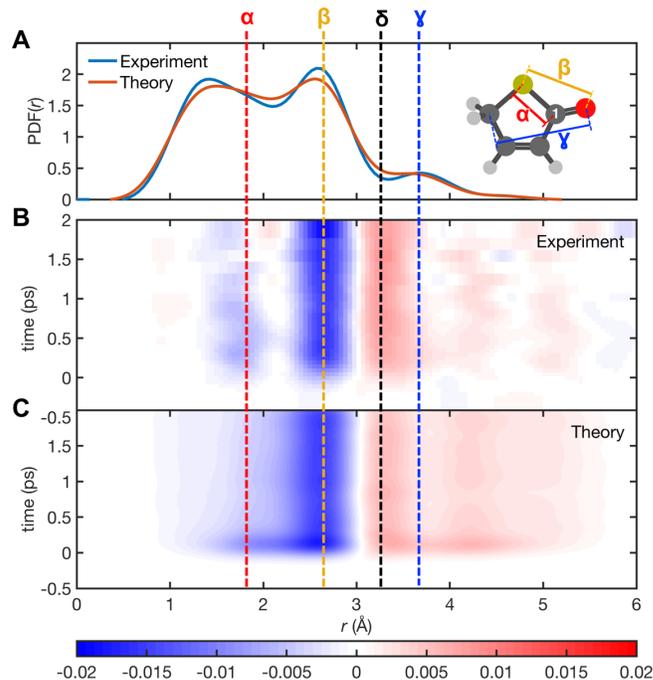

Figure 2. Comparisons between the experimentally-derived and theoretically-predicted PDFs and the $\Delta PDF(r,t))$ maps. (**A**): Experimentally-derived and theoretically-predicted steady-state PDFs for 2(5H)-thiophenone. See SI for the full assignment of the steady-state PDFs for 2(5H)-thiophenone (Fig. S10). (**B**) and (**C**): False-color plots of the experimental and theoretical $\Delta PDF$ as a function of pump-probe time delay. The latter summarizes the dynamics observed for a swarm of 43 trajectories obtained from trajectory surface hopping simulations (SA4-CASSCF(10/8)/6-31G*) for the NAMD from $S_2$ to $S_0$, further continued in $S_0$ with BOMD (UDFT/PBE0/6-31G*). The signal shown in (**C**) was reconstructed from the trajectories using the independent atom model, convolved with a 230-fs full width at half maximum (FWHM) Gaussian function to approximate the IRF. The separations associated with the three strongest features of the static PDF, $\alpha$ (red), $\beta$ (yellow), and $\gamma$ (blue), and an obvious photoinduced feature, $\delta$ (black), are highlighted using vertical dashed lines. The interatomic spacings that contribute most to the $\alpha$, $\beta$ and $\gamma$ features are shown by the color-coded tie-lines overlaid on the optimized structure of ground-state 2(5H)-thiophenone, which also identifies the carbonyl C atom (C1) that features later in the narrative.

We first focus on the theoretically predicted $\Delta I/I(s,t)$ map (Fig. 3A). Each combined (NA+BO)MD simulation returns nuclear coordinates (i.e., a structure) at each time step, from which the $\Delta I/I(s)$ pattern can be calculated (see Methods and the SI). Fig. 3A shows the sum of all such contributions, at each time step. Each instantaneous structure in each simulation can also be assigned to one of three *families* of photoproducts: ring-closed (designating return as an internally excited ring-closed thiophenone); ring-opened (i.e., excited-state biradicals at very early time delays, and the acyclic thioenol or thioaldehyde isomers after reversion to the $S_0$ state); and the proposed episulfide (see Methods and Fig. S4 in the SI for details on the classification).

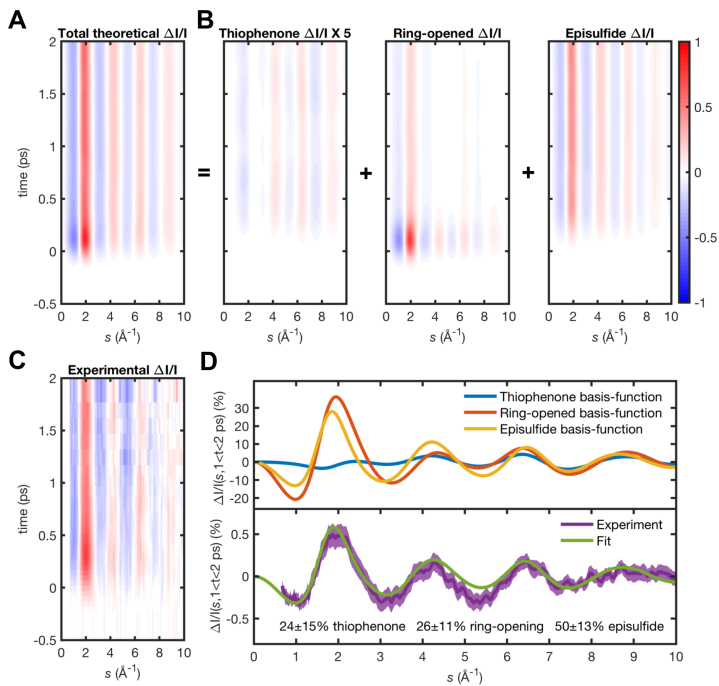

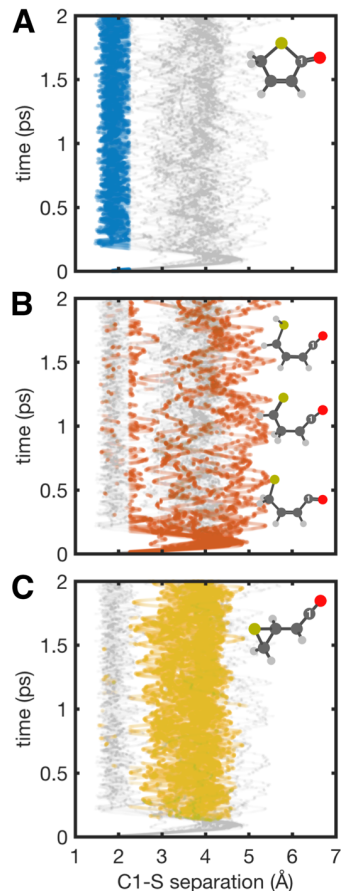

Figure 3. Contributions of the different photoproducts to the $\Delta I/I$ UED signal. (**A**) and (**B**): Theoretical total $\Delta I/I(s,t)$ map and the $\Delta I/I(s,t)$ contributions from each family of photoproducts, each of which has been scaled by the experimental excitation percentage (~3%) and convolved with a 230-fs FWHM Gaussian function to approximate the IRF; (**C**): False-color plot of the measured $\Delta I/I(s,t)$ vs $s$; (**D**), upper panel: Scattering $\Delta I/I(s,t)$ vs $s$ signatures for 2(5H)-thiophenone, ring-opened and episulfide photoproducts, based on the average of the theoretical $\Delta I/I(s,t)$ signals of classified structures extracted from trajectories in the time interval 1 - 2 ps; lower panel: Comparison between the experimental $\Delta I/I(s,t)$ signal and the fit obtained using the theoretical signature for each photoproduct depicted in the upper panel and the stated product branching ratios.

We stress at this point that the simulations show zero formation of the ring-closed 2(3H)-thiophenone within the 2 ps dynamics presented here; ring-closing in the S$_0$ state results solely in re-formation of the parent 2(5H)-isomer. Figs. 4A-4C highlight the very different amplitudes of the C1…S separations associated with instantaneous structures assigned to each photoproduct family. Representative structures from each photoproduct family are displayed in these panels. The total $\Delta I/I(s,t)$ map (Fig. 3A) is simply the sum of the photoproduct specific $\Delta I/I(s,t)$ maps shown in Fig. 3B, weighted according to the relative populations of the photoproduct families.

Inspection of the photoproduct specific and total $\Delta I/I(s,t)$ maps (Figs. 3A-3C) reveals that the ring-opened structures make substantial contributions immediately after photoexcitation, consistent with the initial 2(5H)-thiophenone ring-opening (see the C1…S separation plot in Fig. 4B). They also show that, within the time frame probed, the respective diffraction signatures are largely insensitive to time once all molecules have reverted to the S$_0$ state (*i.e.*, after ~0.5 ps – see SI for an extended discussion). Fig. 3D (upper panel) shows the 'average' $\Delta I/I(s)$ signatures derived from the (NA+BO)MD simulations for each photoproduct family during the period +1.0 ≤ *t* ≤ +2.0 ps, which we can use as time-independent basis functions (see Fig. S6 in the SI) to decompose the total $\Delta I/I(s,t)$ maps. This is a key aspect of the present analysis: using these basis functions to fit the experimental $\Delta I/I(s,t)$ signal integrated over the period +1.0 ≤ *t* ≤ +2.0 ps provides clear evidence that the episulfide is formed upon UV-irradiation of 2(5H)-thiophenone (Fig. 3D, lower panel).

Figure 4. Calculated time-varying C1…S separation for the full swarm of trajectories (gray) and, respectively, 2(5H)-thiophenone (**A**, blue), ring-opened molecules (**B**, orange) and episulfide (**C**, yellow).

The best reproduction of the features observed in the raw data between 1.0 ≤ *s* ≤ 4.0 Å$^{-1}$ (the region of reciprocal space that offers the best signal to noise ratio) is obtained by including a substantial contribution from the basis function for episulfide (see SI for further validations of this procedure and the potential limitations of treating the data with a high band pass filter). Further, the distinctive nature of the UED signal allows estimation of the relative population of the episulfide photoproduct: 50 ± 13 % over the 1-2 ps time window. Strictly, this latter decomposition returns the time-dependent fractional contribution of each basis signature to the experimental $\Delta I/I(s,t)$ data. These ratios will correspond to photoproduct population ratios provided that the independent atom model is valid, *i.e.*, that each isomer of a common ground-state species (here the neutral C$_4$H$_4$SO species) shows the same elastic diffraction intensity.[37, 40-42]

The present study also serves to emphasize the importance of extracting such theoretical basis functions from geometries representative of the athermal distribution of the photoproducts, *not* simply from their ground-state optimized geometries or from some assumed thermalized distribution. Fig. S15 in the SI illustrates the very different static PDF(*r*) and *sM*(*s*) profiles predicted for such distributions of 2(5H)-thiophenone molecules. The quantitative analysis of the UED signal provided here requires great care in the processing of the data, as is also detailed in the SI (Figs. S6-S8). Overall, the combination of UED and theory proves episulfide formation within a few hundreds of fs following 266-nm photoexcitation of 2(5H)-thiophenone.

Extracting time-resolved photoproduct populations

The foregoing analysis returned a relative population of episulfide photoproducts in the $+1.0 \leq t \leq +2.0$ ps time window, but the fitting strategy combining theory and experimental UED signals allows us to go a step further: performing an analogous fit to UED data recorded at each experimental time delay reveals the *time dependence* of the different photoproduct populations.

The (NA+BO)MD-predicted, time-dependent relative populations are shown by dotted curves in Fig. 5A and, again, as solid curves after convolving with a 230-fs FWHM Gaussian function used as an approximate IRF. Figure 5B shows the results of a similar decomposition of the UED-derived total $\Delta I/I(s,t)$ map shown in Fig. 3C. Experiment and theory are in excellent agreement with regard to the ultrafast disappearance of the parent 2(5H)-thiophenone molecule (blue traces in Fig. 5) and the rise of a ring-opened structure (orange traces) following photoexcitation – in this case a biradical photoproduct emerging from the ring-opening of 2(5H)-thiophenone. Theory and experiment further agree on the timing of the birth of episulfide (yellow traces in Fig. 5), taking place ~200 fs after the appearance of the ring-opened photoproduct.

While theory and experiment agree on the photoproduct relative populations at 2 ps, theory predicts a faster growth of episulfide population. This difference can be explained by the way the NAMD is initialized at $t = 0$, where the ground-state molecular wavefunction is perfectly projected onto $S_2$, mimicking the photoexcitation that would be obtained with a δ-pulse.[43] This initialization represents the formation of a perfectly coherent nuclear wavepacket in $S_2$ – a molecular state that only approximates the experimentally-induced molecular state obtained by the progressive population of the excited state with a 67-fs laser pulse. We also note that the (NA+BO)MD employs a swarm of classical, independent trajectories whose evolution tends to appear more (classically) coherent.

Notwithstanding the foregoing caveat, the level of agreement between experiment and theory is very satisfying. Both show that photoexcitation to the $S_2$ state drives immediate C1–S bond extension (manifested as loss of 2(5H)-thiophenone and a decline in the blue trace in Fig. 5), leading to ring-opening (the early time rise in the orange trace in Fig. 5), which enables reversion to the $S_0$ state.[17] These conclusions are entirely consistent with those of the previously reported TRPES study of UV photoexcited 2(5H)-thiophenone molecules,[17] with the critical difference that the UED measurements presented here additionally allow us to (i) identify the episulfide unequivocally and (ii) extract time-dependent relative populations for the different photoproduct families.

Dynamics govern the product population distributions

The present UED measurements and complementary modeling also provide a wealth of hitherto inaccessible dynamical insights. Once back in the $S_0$ state, the ring-opened biradical species either convert to ring-closed parent 2(5H)-thiophenone molecules (revealed by the rise after 250 fs in the blue trace in Fig. 5) or form ring-opened and episulfide photoproducts (shown by, respectively, the orange and yellow traces in Fig. 5). We reiterate that the molecular simulation shows that ring-closing results exclusively in re-formation of the 2(5H)-parent isomer. As also noted above, the UED signal analysis cannot distinguish between the three acyclic (ring-opened) photoproducts, but information on the appearance of these molecules can be extracted from the (NA+BO)MD dynamics. The simulations (Fig. S14) show ground-state 2-(2-sulfanethyl)ketene and 2-thioxoethylketene (Scheme 1) appearing within 100 fs of the photoexcitation process, with respective populations fluctuating around ~10% of all photoproducts within 200 fs. The simulations also illustrate the rapidity (sub-picosecond timescale) of the interconversion between the different highly internally excited ring-opened photoproduct isomers by H atom transfer between adjacent heavy atoms. The decline of the ring-opened biradical population exhibits two timescales: an initial fast decline within 250 fs, followed by a further more gradual decline to below 10% of the total photoproduct population within the 2 ps of the present dynamics simulations (see SI).

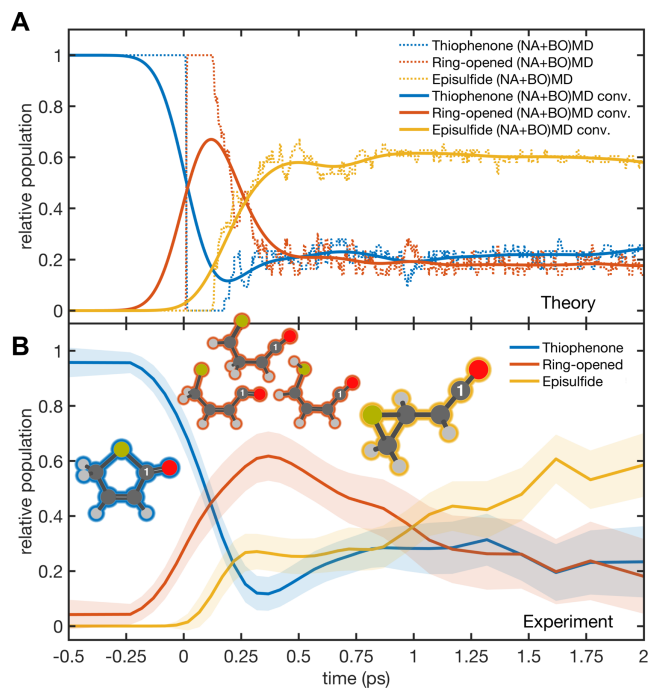

Figure 5. Time-resolved relative populations for the three families of photoproducts. (**A**): Time-resolved relative populations of product isomers obtained directly from the combined (NA+BO)MD simulations: 2(5H)-thiophenone (blue), ring-open forms (orange) and episulfide (yellow). The classification used to unravel these populations is discussed in the SI. The dotted lines show the populations as obtained from the dynamic calculations, while the solid lines are the same data convoluted by a function representing the IRF. (**B**): Time-dependent relative populations of products extracted from the experimental UED signal.

It is also important to note that energy and momentum conservation dictate that all these photoproducts must carry high levels of vibrational excitation, the distribution of which amongst the available normal modes will be determined by the dynamics of the initial bond fission. Such athermal early time nuclear motions are increasingly becoming recognized as a potential trigger for unexpected (*i.e.*, non-statistical) ground-state reactivity.[9, 44-48] The (NA+BO)MD simulations highlight the fluxional nature of these highly internally excited photoproducts, which inter-convert between the various structural families on a sub-ps timescale (as can be seen from the all-trajectory C1…S separation vs time plots shown in Fig. 4). Thus, it is important to appreciate that, though the UED (and (NA+BO)MD) data and their analyses provide product branching ratios, the photoproduct identities are not frozen by the end of the time window sampled in the present study. Any one photoproduct molecule carries more than sufficient internal energy to be able to inter-convert amongst other photoproduct families.

Comparisons with the respective product populations for a canonical distribution of ground state molecules maintained at a temperature corresponding to $E \sim 4.65$ eV ($T \sim 2200$ K) are also instructive. A simple Boltzmann calculation based on the respective ground state electronic energies returns the population ratios thiophenone : ring-opened : episulfide ~ 99.87 : 0.08 : 0.05%. Some 31% of the thiophenone species would be predicted to be in the form of the 2(3H)- isomer given the relative ground state energies shown in Scheme 1. The observed predominance of episulfide photoproducts and the complete absence of 2(3H)-thiophenone ring-closed products returned by the simulations emphasizes that the early-time photoproduct population distributions are dynamically determined. As noted before, anharmonicity will encourage a gradual 'thermalization' of internal energy amongst all normal modes of the 'hot' photoproducts and an evolution towards a more statistical product population distribution over much longer time scales. However, as also noted previously,[17] the total available internal energy is sufficient to enable unimolecular decay of the acyclic photoproducts (*e.g.*, CO loss) at later times.

The agreement between theory and experiment confirms the exotic episulfide as a major photoproduct, with a relative population of 50 ± 13 % at early times. Episulfide product formation explains the different photoinduced reductions in the features labelled α and β in Fig. 2: the O….S separation in the episulfide (and the ring-opened photoproducts) is much larger than that in the parent 2(5H)-thiophenone, so all such transformations lead to a reduction in the peak labelled β. But the episulfide photoproducts contain C–S bond separations similar to those in 2(5H)-thiophenone. This, plus the fact that one C–S bond survives in the initial ring-opening, explains the lesser reduction in the intensity of the feature labelled α. Even the β peak is likely to be less reduced than might be expected based on the zero-order assumption that it reports just on the O…S separation, since the longer C…O separation in the C=C=O group within the photoproducts will also contribute to elastic diffraction signal at $r \sim 2.5$ Å. All photoproducts contain C…O pairs with similar separations to that in the 2(5H)-thiophenone precursor – accounting for the apparent insensitivity of the feature labelled γ to photoexcitation. Reference to Fig. 4C shows that the C1…S separation in the episulfide product is a major contributor to the photoinduced feature labelled δ in Fig. 2.

## CONCLUSION

This study illustrates how a combination of *ab initio* (nonadiabatic and adiabatic) molecular dynamics simulations and contemporary MeV-UED probe techniques allow (i) unequivocal identification of the episulfide isomer following UV excitation of 2(5H)-thiophenone and (ii) determination of time-dependent relative populations of the different families of photoproduct isomers at early times following nonadiabatic coupling back to the ground state potential. As noted at the outset, reversion to the ground electronic state is a common fate for molecules following photoexcitation. This demonstration study highlights the importance of dynamics in determining the relative populations of photoproducts and the rapid (sub-picosecond timescale) interconversion between the different highly internally excited acyclic photoproduct isomers enabled by transferring an H atom between neighboring heavy atoms. It points the way to a wealth of future studies designed to identify all major products from such photoinduced athermal ground-state chemistry and to determine their relative populations and how these evolve at early times – a spread of insights currently offered by few (if any) other ultrafast probe methods.

## ASSOCIATED CONTENT

### Supporting Information

The Supporting Information contains a detailed description of the experimental setup, data processing, error estimation, full computational details, fitting procedures and benchmarking. (PDF)
The raw UED dataset (15mJ) and trajectories used in this work are available at the following link: doi.org/10.5281/zenodo.10045070.

The Supporting Information is available free of charge on the ACS Publications website.

## AUTHOR INFORMATION

### Corresponding Authors
* MNRA: mike.ashfold@bristol.ac.uk
* DR: rolles@phys.ksu.edu
* BFEC : basile.curchod@bristol.ac.uk

### Present Addresses
$^§$Center of Basic Molecular Science, Department of Chemistry, Tsinghua University, Beijing, China.
$^#$SRI International, Boulder, CO 80302, USA.
$^=$Vescent Photonics, Golden, CO 80401, USA.

### Author Contributions
SP and DR conceived the experiment with input from MC, JPFN, and TJAW. ARA, MFL, RJGF, NG, MH, DL, ARe, XS, XJW, MRW, SPW, TJAW, and JY operated the experiment with remote participation of JPFN, SP, SB, RB, KB, MC, JC, BE, CSH, DMPH, RAI, SBM, ARo, ARu, SKS, ASV, KW, YX, MNRA, and DR. LMI, JPFN, and BFEC conceived the theoretical methodology for the analysis of the experimental data with input from DR, MNRA, and SP. LMI and BFEC performed the *ab initio* simulations, and JPFN calculated the theoretical scattering signal from the *ab initio* trajectories. SP and JPFN analyzed the experimental data with input from DR, MNRA, LMI, and BFEC. SP, JPFN, DR, MNRA, LMI, and BFEC interpreted the data and wrote the manuscript with input from all the authors.
†These authors contributed equally.

## FUNDING SOURCES

The SLAC MeV UED facility is supported in part by the US Department of Energy, Office of Basic Energy Sciences, SUF Division Accelerator & Detector R&D program, the Linac Coherent Light Source Facility, and SLAC under contract nos. DE-AC02-05-CH11231 and DE-AC02-76SF00515. Other authors are funded through: National Science Foundation grant PHYS1753324 (AV, DR); Chemical Sciences, Geosciences, and Biosciences Division, Office of Basic Energy Sciences, Office of Science, US Department of Energy (TJAW) under grant no. DE-FG02-86ER13491 (SP, AR), DE-SC0019451 (KB), DE-SC0020276 (SB), DE-SC0017995 (NG), and DE-SC0020276 (JPFN, SS, MC); Engineering and Physical Sciences Research Council grant no.

EP/L005913/1 (MNRA), EP/V026690/1 (BFEC), EP/X026973/1 (BFEC), and EP/R513039/1 (LMI); ANR Q-DeLight project, Grant No. ANR-20-CE29-0014 of the French Agence Nationale de la Recherche (LMI); Australian Research Council grant no. DE200100549 (CSH); European Union Horizon 2020 research and innovation programme grant no. 803718 (BFEC); Science and Technology Facilities Council (DMPH).

**ACKNOWLEDGMENT**

We thank the technical and scientific team at the SLAC MeV UED facility for their excellent communications and support before and during the beamtime, which was performed as a remote-user experiment due to the COVID pandemic. The theoretical and computational part of this work used the facilities of the Hamilton HPC Service of Durham University. DR is thankful to LCLS and the Stanford PULSE Institute for their hospitality and financial support during a sabbatical.

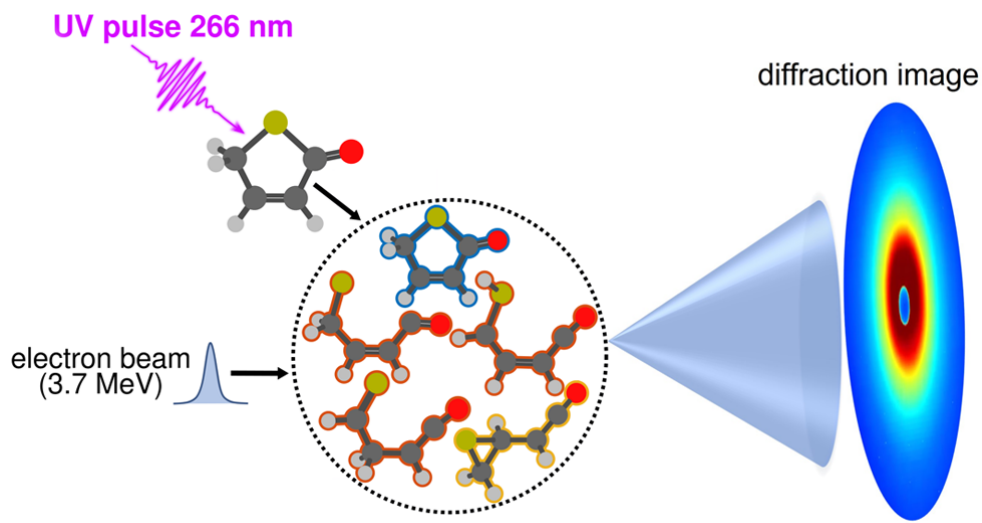

Supporting Information for

# Monitoring the evolution of relative product populations at early times during a photochemical reaction

Joao Pedro Figueira Nunes[†1], Lea Maria Ibele[†2], Shashank Pathak[†3], Andrew R. Attar[4=], Surjendu Bhattacharyya[3], Rebecca Boll[5], Kurtis Borne[3], Martin Centurion[1], Benjamin Erk[6], Ming-Fu Lin[4], Ruaridh J.G. Forbes[4], Nate Goff[7], Christopher S. Hansen[8], Matthias Hoffmann[4], David M.P. Holland[9], Rebecca A. Ingle[10], Duan Luo[4], Sri Bhavya Muvva[1], Alex Reid[4], Arnaud Rouzée[11], Artem Rudenko[3], Sajib Kumar Saha[1], Xiaozhe Shen[4], Anbu Selvam Venkatachalam[3], Xijie Wang[4], Matt R. Ware[4#], Stephen P. Weathersby[4], Kyle Wilkin[1], Thomas J.A. Wolf[4,12], Yanwei Xiong[1], Jie Yang[4,$], Michael N. R. Ashfold*[13], Daniel Rolles*[3], Basile F. E. Curchod*[13]

[1]*University of Nebraska–Lincoln, Lincoln, NE, USA;* [2]*CNRS, Institut de Chimie Physique UMR8000, Université Paris-Saclay, Orsay, France;* [3]*J.R. Macdonald Laboratory, Physics Department, Kansas State University, Manhattan, KS, USA;* [4]*SLAC National Accelerator Laboratory, Menlo Park, CA, USA;* [5]*European XFEL, Schenefeld, Germany;* [6]*Deutsches Elektronen Synchrotron DESY, Hamburg, Germany;* [7]*Brown University, Providence, RI, USA;* [8]*School of Chemistry, University of New South Wales, Sydney, NSW, Australia;* [9]*Daresbury Laboratory, Warrington, UK;* [10]*Department of Chemistry, University College London, London UK,* [11]*Max Born Institute, Berlin, Germany;* [12]*Stanford PULSE Institute, SLAC National Accelerator Laboratory, Menlo Park, CA, USA;* [13]*School of Chemistry, University of Bristol, Bristol, UK*
[$]*Present address: Center of Basic Molecular Science, Department of Chemistry, Tsinghua University, Beijing, China.*
[#]*Present address: SRI International, Boulder, CO 80302, USA.*
[=]*Present address: Vescent Photonics, Golden, CO 80401, USA.*

†These authors contributed equally to this work.
*mike.ashfold@bristol.ac.uk; rolles@phys.ksu.edu; basile.curchod@bristol.ac.uk

**Material and Methods**

*Experiment*

*Experimental data processing:*

Two-dimensional diffraction patterns recorded at the EMCCD detector were processed into one-dimensional scattering intensity curves, *I(s)*, using the workflow listed below:

1. Outliers removal - Pixels with intensities 4 standard-deviations above the mean of identical pixels, *i.e.* same time delay and detector coordinate, were removed.
2. Data averaging - Diffraction patterns acquired at the same time delay were averaged together.
3. Background removal - An estimated background frame generated by linearly interpolating between the intensities of the four corners of the average image frame was subtracted from all diffraction image frames.
4. Detector mask - Masks were applied to all areas of the detector which do not image the phosphor screen and therefore do not contain scattering information. These areas include the edges of the image and the hole at the center of the detector, through which the transmitted electron beam passes.
5. Diffraction center assessment - The center of each diffraction pattern was determined by fitting circles to pixels with identical intensities (isolines).
6. Radial outlier removal - Pixels with intensities over 3 standard-deviations above the mean of pixels at the same radial distance from the diffraction center were removed.
7. Data reduction - Two-dimensional diffraction images frames were radially averaged around the diffraction pattern center to produce one-dimensional scattering intensity curves.
8. Data normalization - Scattering intensity curves were normalized based on the mean counts across the detected momentum transfer range: $0.67 < s < 12$ Å$^{-1}$.
9. Baseline subtraction - A power function ($a \times x^b$) is fitted to, and subtracted from, all one-dimensional scattering intensity curves.

*Generating pair distribution functions from the experimental data:*

Scattering intensities are measured at the detector as a function of the momentum transfer vector, *s*, defined as:

$$s = \frac{4\pi}{\lambda} sin(\frac{\theta}{2}) \qquad (eq.\ S1)$$

where $\lambda$ is the de Broglie wavelength of the incident electrons and $\theta$ is the angle between the incident and scattered electrons. Under the independent atom model (IAM) approximation, the total scattering intensity, $I(s)$, can be decomposed into molecular and atomic scattering contributions:

$$I(s) = I_{at}(s) + I_{mol}(s) \tag{eq. S2}$$

The atomic scattering term, $I_{at}(s)$, does not contain structure information and can be calculated easily, provided the empirical formula of the target molecule is known. For a molecule containing $N$ atoms, the atomic scattering term is expressed as the sum of each atomic differential cross-section (each being equal to the modulus squared of the elastic scattering amplitude $f_i(s)$):

$$I_{at}(s) = \sum_{i=1}^{N} |f_i(s)|^2 \tag{eq. S3}$$

The scattering amplitudes for a 3.7 MeV incident electron were calculated using the ELSEPA code.[2] The molecular scattering term, $I_{mol}(s)$, encodes information on the internuclear distances in the target molecules. For an isotropic sample, $I_{mol}(s)$ is expressed as the sum of interference terms for all possible atom pairs:

$$I_{mol}(s) = \sum_{i=1}^{N} \sum_{j \neq i}^{N} |f_i(s)||f_j(s)| \frac{\sin(sr_{ij})}{sr_{ij}} \tag{eq. S4}$$

where $f_i(s)$ and $f_j(s)$ are the elastic scattering amplitudes for the $i^{th}$ and the $j^{th}$ atoms, respectively; and $r_{ij}$ is the internuclear distance between the $i^{th}$ and $j^{th}$ atoms.

Diffraction data are typically presented in the form of modified scattering intensities which enhance the oscillations in the $I_{mol}$ term and suppress the rapid drop in scattering intensity as a function of $s$ imparted by the $s^{-2}$ scaling in the elastic scattering amplitude. The modified scattering intensity, $sM(s)$, is defined as:

$$sM(s) = \frac{I_{mol}(s)}{I_{at}(s)} s \tag{eq. S5}$$

A method, developed by Ihee et al.,[3] can be used to calculate experimental modified scattering intensities, $sM_{exp}(s)$, as the $I_{mol}$ and $I_{at}$ cannot be separated experimentally:

$$sM_{exp}(s) = \frac{I_{exp}(s) - I_{bkg}(s)}{I_{at}(s)} s \tag{eq. S6}$$

where $I_{exp}(s)$ is the experimentally measured scattering intensity and $I_{bkg}(s)$ is an estimate of the instrument-specific background and atomic scattering contributions. In this work, the $I_{bkg}(s)$ term was approximated by fitting a sum of exponents to the zero-crossing of the theoretical $I_{mol}(s)$ term for 2(5H)-thiophenone.

The $sM(s)$ curve can be approximated to a sum of sine waves, arising from all internuclear distances in the target molecule. Therefore, the $sM(s)$ curve can be decomposed into a pair-distribution function (PDF) of all contributing interatomic distances using the following sine transform:

$$PDF(r) = \int_0^{s_{max}} sM(s)\sin(sr)e^{-ks^2}ds \quad \text{(eq. S7)}$$

where $s_{max}$ is the maximum transfer in the diffraction with adequate signal-to-noise ratio, $r$ is the internuclear distance between atom pairs, and $k$ is a damping factor used to suppress the high $s$ contribution smoothly to zero. A damping factor of 0.03 was used in all static PDF calculations. Prior to the sine transform of $sM_{exp}(s)$ curves, low-scattering angle data ($s < 0.7$ Å$^{-1}$) obscured by the hole in the detector was filled in using a linear extrapolation to $s = 0$ in order to minimize the impact of edge artifacts in the experimental PDF.

*Generating difference signals from experimental data:*

The analysis of time-resolved experimental UED data is based on the difference-diffraction method.[4] This approach suppresses both contributions from molecules not excited by the pump laser and instrument-specific background contributions, thus enhancing the signal arising from photoexcited molecules undergoing structural rearrangements. Our analysis employs a robust representation of the difference-diffraction signal: the fractional change signal, henceforth referred to as $\Delta I/I(s,t)$:

$$\Delta I/I(s,t) = \frac{I(s,t) - I(s,t<0)}{I(s,t<0)} \quad \text{(eq. S8)}$$

where $I(s,t<0)$ is the reference diffraction signal taken before the arrival of the pump pulse, and $I(s,t)$ is the diffraction intensity recorded at pump-probe delay $t$. Background contributions unaccounted for by the difference-diffraction method are removed from the experimental difference-signal by the fitting and subtracting of a low-order polynomial from the $\Delta I/I(s,t)$.

The experimental time-dependent difference pair distribution functions, $\Delta PDF(r,t)$, shown in Fig. 2 of the main text, were calculated by applying the sine-transform of time-dependent difference-modified scattering curves, $\Delta sM_{exp}(s,t)$:

$$\Delta sM_{exp}(s,t) = \frac{I(s,t) - I(s,t<0)}{I_{at}(s)}s \quad \text{(eq. S9)}$$

$$\Delta PDF(r,t) = \int_0^{s_{max}} \Delta sM_{exp}(s,t)\sin(sr)e^{-ks^2}ds \quad \text{(eq. S10)}$$

A damping factor of 0.03 is used in all PDF and $\Delta$PDF calculations and missing low-scattering angle data was filled in by linear extrapolation to zero.

*Detector size to momentum transfer vector calibration:*

The conversion between the detector pixel size and the momentum transfer vector, $s$, was calibrated using the known positions of Bragg reflection from a bismuth telluride ($Bi_2Te_3$) single crystal sample measured at the beginning of the experiment. The value of this conversion was then optimized for each dataset by comparing the theoretical and experimental static (no optical pump) scattering signatures for 2(5H)-thiophenone.

*Time-zero determination:*

During the experiment, the position of the time-zero was estimated based on the Debye-Waller profile of the diffraction signal of an optically pumped single crystal of silicon. This position was refined during the data analysis process to represent the half-maximum of the absolute percentage change of the difference-diffraction signal.

*Error estimation of the experimental signal:*

The statistical uncertainty of the experimental signal was estimated using a standard bootstrapping analysis. The UED dataset, which consisted of a pool of 178 unique scans, was randomly resampled with replacement 150 times to produce 150 bootstrapped datasets. Each bootstrapped dataset was analyzed separately, thereby enabling the mean and standard deviation for all relevant analysis outputs to be evaluated. Fig. S2 shows the uncertainty of the UED measurement represented as one standard deviation of the calculated difference-diffraction signal, $\Delta I/I(s,t)$, across the bootstrapped datasets. Note that the estimated uncertainty of the measurement is substantially smaller than the amplitude of the signal across all positive time delays. Saturation of bootstrapped uncertainty reflected in the difference-diffraction standard deviation and standard error was observed after 150 bootstrapped datasets (see Fig. S3).

*Assessment of the presence of photoionization in the UED signal:*

The plasma lensing effect, first reported by Dantus and Zewail,[5] was used to assess the presence of photoionization in the probed sample volume. Briefly, the plasma field generated by the separation of charges during an ionization event induces a deflection in the incident electron beam.[6] This phenomenon is expressed in the UED signal as a strong persistent difference signal at low-scattering angle ($s < 1$ Å$^{-1}$) and is accompanied typically by a lensing of the undiffracted electron beam ($I_0$ signal).[7] The fluence-dependence of the difference signal at $s < 1$ Å$^{-1}$, depicted in Figure S13A, shows no appreciable change in signal levels at low-scattering angles for pump energies below 25 μJ. Therefore, we conclude that photoionization is unlikely to be present in the 15 μJ dataset discussed in this manuscript. This observation is corroborated further by the lack of electron beam lensing or deflection shown in Figure S13D-F.

*Theory*

*Generation of theoretical scattering signals:*

The theoretical static PDF($r$) and time-dependent $\Delta I/I(s,t)$ and $\Delta PDF(r,t)$ signals in the main text were calculated in accordance to the independent atom model (IAM) [19, 20] using the nuclear configurations captured along the 43 (NA+BO)MD trajectories and eqs. 7, 8 and 10, respectively. The BOMD trajectories were refined with a monotonic time step of 0.5 fs (accounting for the difference in time step: 0.5 fs for NAMD and 0.1 fs for BOMD).

*Retrieval of photoproduct relative populations:*

The relative populations for the photoproducts were retrieved directly from the UED signal by fitting a linear combination of basis functions selected to reflect the average scattering signatures of photoproducts to the experimental difference-diffraction signal, $\Delta I/I(s,t)$. This approach allowed the relative abundances, *i.e.,* relative populations of 2(5H)-thiophenone, ring-opened and episulfide products to be determined experimentally and their evolution across the experimental time window mapped. The fitting of photoproducts was carried out in reciprocal space as it provides a more robust experimental signal from which quantitative information can be extracted. Although more intuitive, the analysis of real-space signals generated by the sine-transform of difference-diffraction signals requires the implementation of artifact mitigation strategies, such as the filling of missing data at low-scattering angles and the damping of high-scattering angle contributions, both of which impact the breath and amplitude of the calculated $\Delta PDF$. For this reason, the real-space features are considered to offer a less quantitative signal from which to determine information regarding relative abundances of photoproducts.

*Basis function selection:*

The three photoproduct basis functions used to fit the experimental difference-diffraction signal were obtained by averaging the theoretical $\Delta I/I(s,t)$ signals for 2(5H)-thiophenone, ring-opened and episulfide products in the range $1 \leq t \leq 2$ ps. The temporal evolutions of the two strongest scattering features in the theoretical $\Delta I/I(s,t)$ functions for these three photoproducts, depicted in Figs. S6B and S6C, plateau after ~ 1 ps. This plateau reflects the settling of photoproduct geometries into discrete ensembles with a breadth of conformations that adequately represents each photoproduct classification. For this reason, the average signal in the range $1 \leq t \leq 2$ ps was selected as an adequate time-independent basis function with which to fit the experimental signal. However, a consequence of using time-independent basis functions based on the average scattering signals of vibrationally hot photoproducts in the ground state is that they are unlikely to optimally capture all structural changes taking place in the early times following photoexcitation, when the molecule is in an excited electronic state.

*Fitting routine:*

A global search algorithm implemented in Matlab was used to minimize the root-mean-square error (RMSE) between the theoretical $\Delta I/I(s,t)$ signal calculated from a linear combination of photoproduct basis functions and the experimental $\Delta I/I(s,t)$ signal for each time delay visited in the UED experiment. The fits were performed on the raw experimental signal without the application of high or low pass filters. This avoids the biasing of the fit by an arbitrary selection of high and/or low pass cutoff frequency. During the fitting routine, the theoretical signal is scaled to reflect the percentage of molecules in the experimentally probed volume that were optically excited by the UV pulse and therefore contribute to the experimental difference-diffraction signal. This scaling factor, which is constant across the entire dataset, was determined by fitting the average experimental signal between 1 and 2 ps to a scaled linear combination of basis functions. The results of this fit, which are shown in the bottom panel of Fig. 3D, indicate that 3% of the molecules in the probed volume are photoexcited (scaling factor of 0.03). To remove biases introduced by the selection of the momentum transfer range included in the fit, fitting routines were repeated 8 times with randomized starting and end points between 0.65-2 and 7-10 Å, respectively. Moreover, the fitting of both the average experimental signal between 1 and 2 ps and the fitting of the experimental signal for unique time delays was repeated for each of the 150 bootstrapped datasets. The reported relative population uncertainties are, therefore, a reflection of both the variance of results across different momentum transfer range selections and the inherent variability of bootstrapped datasets. The time-evolving electron diffraction signals reported here could, in principle, surely be reproduced by several combinations of bond distances/angles. It is important to stress that the present interpretation is specifically based on the results of *ab initio* nonadiabatic and adiabatic molecular dynamics simulations of the excited- and ground-state dynamics of 2(5H)-thiophenone following photoexcitation at 266 nm. The exact same simulations were used to interpret and reproduce the results from TRPES studies of 2(5H)-thiophenone following excitation under very similar conditions.[11]

*Benchmarking of the relative population retrieval methodology:*

The performance of the quantitative retrieval of photoproduct relative populations was assessed by applying the fitting routine described above to the total theoretical difference-diffraction signal. The results of this benchmark fit show that the proposed methodology is able to retrieve the relative populations of the three photoproducts identified by classification II. Good quantitative agreement was found between the true relative populations obtained from the (NA+BO)MD simulations (fraction of trajectories showing a given photoproduct at a given time), depicted in Fig. S7A, and those retrieved by fitting the temporally convolved theoretical UED signal, shown in Fig. S7B. As discussed above, the basis functions appear to map less well to the structural rearrangements observed at early pump-probe delays, when the molecule is in an excited electronic state, as seen by the slight discrepancies observed between true theoretical and retrieved relative populations obtained for the 0 – 350 fs time window.

The impact of basis function selection on the goodness of fit was investigated by inspecting the RMSE obtained for a series of fits using different combinations of three photoproduct basis functions. For added flexibility, classification I was used to generate the three photoproduct basis functions, as it yields two additional basis functions corresponding to the ring-opened photoproducts P1 and P2 (the nomenclature for the photoproduct is defined in Figure S4). Figure S8, which depicts the fit of the average experimental signal between 1 and 2 ps using three different combinations of basis functions, shows that the inclusion of a basis function for episulfide results in a better overall fit and allows the close reproduction of the experimental signal in the $1 < s < 4$ Å$^{-1}$ region. It is important to note that this region of momentum transfer range offers the best signal to noise ratio and therefore can be used to assess the impact of basis function selection on the goodness of fit. We also emphasize that no high band pass filter is applied to the raw data to avoid arbitrarily selecting a cut-off and the potential artifacts that such a procedure can induce. The results of these fits, which are summarized in Table S1, show that the inclusion of an episulfide basis function substantially improves the goodness of fit, *i.e.*, lowers the fit RMSE below $8 \cdot 10^{-4}$, for all combinations of basis functions tested. Therefore, episulfide geometries are crucial to the appropriate modeling of scattering signals which arise from the ensemble of photoproducts produced during the UED experiment. Moreover, the results of fits carried out using 3 and 4 photoproduct basis functions, summarized in Table S2, show that the episulfide relative population is insensitive to the exact number and nature of any additional basis functions included in the fit. Therefore, the relative population for episulfide does not act as a sink of residuals in the fitting routines but is a robust fitting parameter.

**Supporting Figures and Tables**

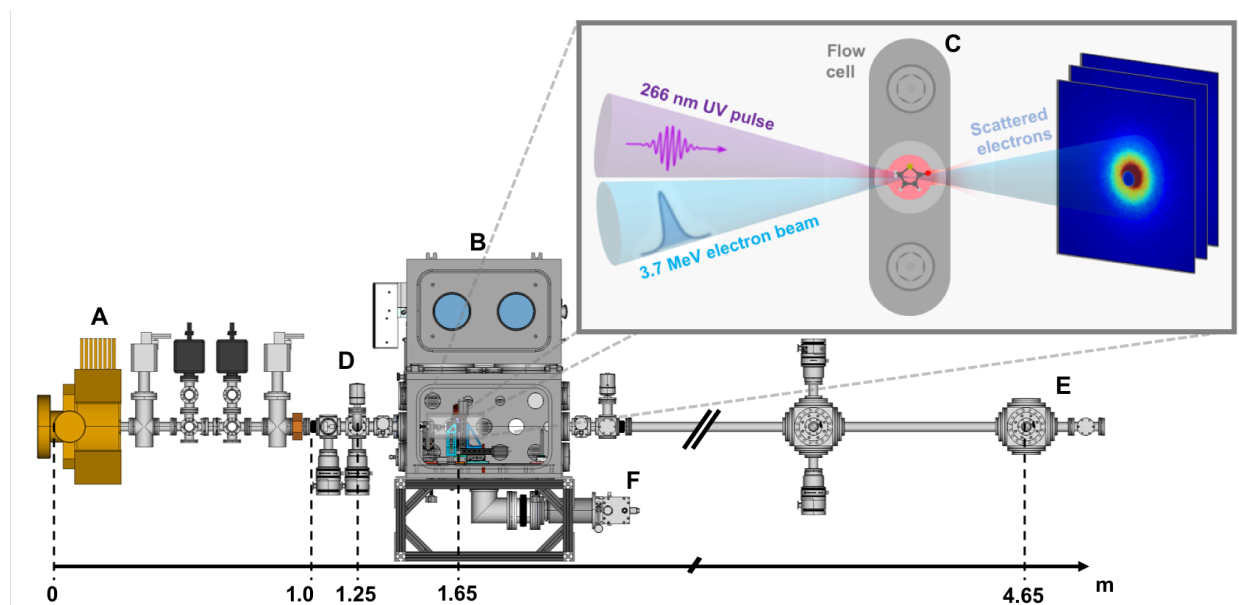

**Fig. S1: Schematic of experimental setup.** At the SLAC National Accelerator Laboratory UED instrument, 3.7 MeV electron bunches generated by a photocathode radio frequency electron gun (A) are accelerated towards an experimental chamber (B) where they impinge upon a volume of 2(5H)-thiophenone gas inside a flow cell (C). The sample volume is optically pumped by a UV laser pulse coupled into the experimental chamber by an in-vacuum holey mirror (D). The scattered electrons are collected 3 meters downstream of the pump-probe overlap region by a scintillator coupled CCD detector (E). Exhausted 2(5H)-thiophenone gas is condensed in a cryogenically cooled high surface area cold trap (F). The inset panel shows a schematic representation of the UED interaction volume and resulting diffraction pattern.

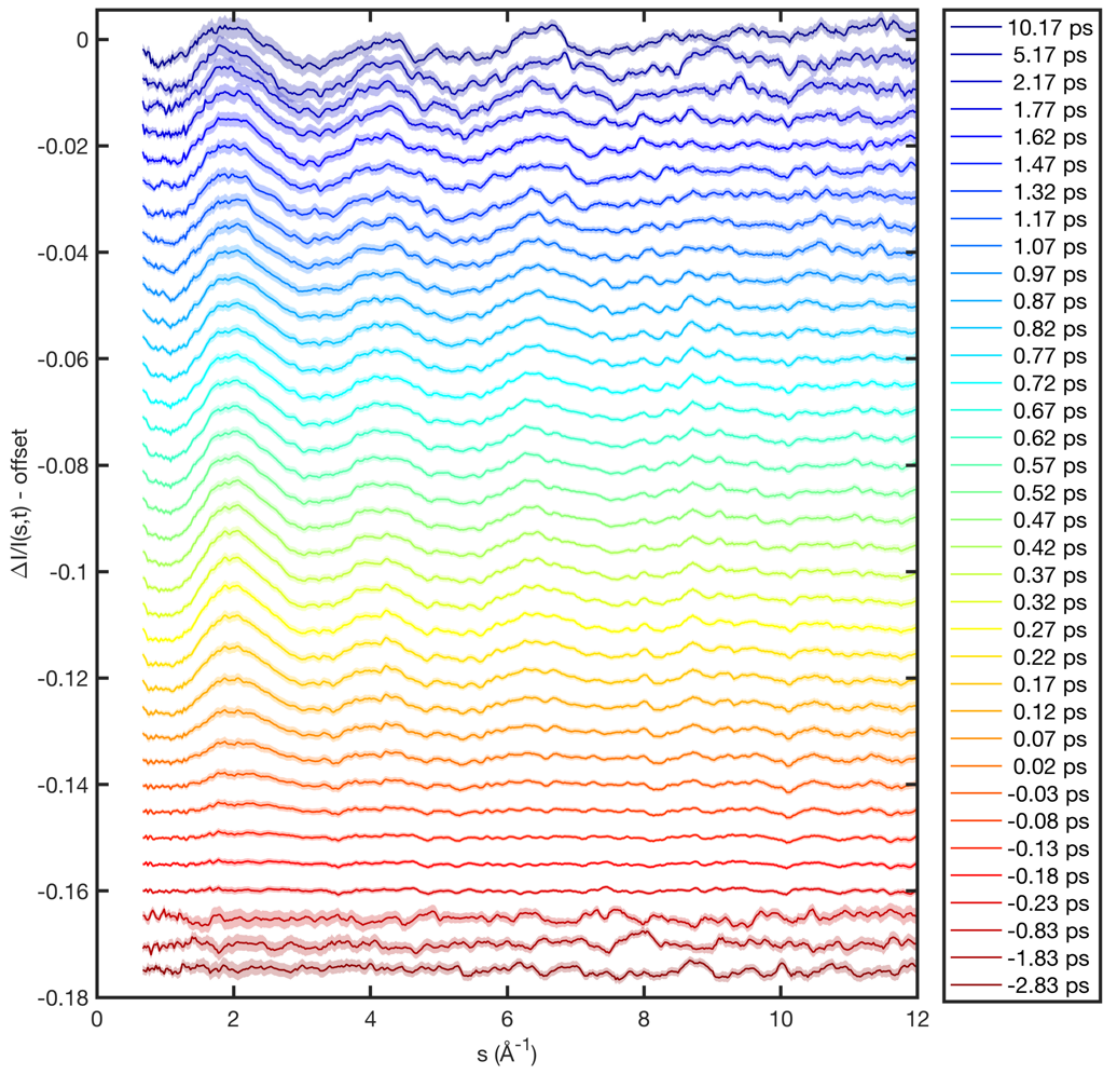

**Fig. S2: Experimental signal uncertainty.** Experimental difference-diffraction signal, $\Delta I/I(s,t)$, for all 37 time delays visited in the UED experiment. Positive and negative delays represent, respectively, the UV pump pulse arriving before and after the UED probe pulse. The shaded areas reflect one standard deviation across the 150 bootstrapped datasets.

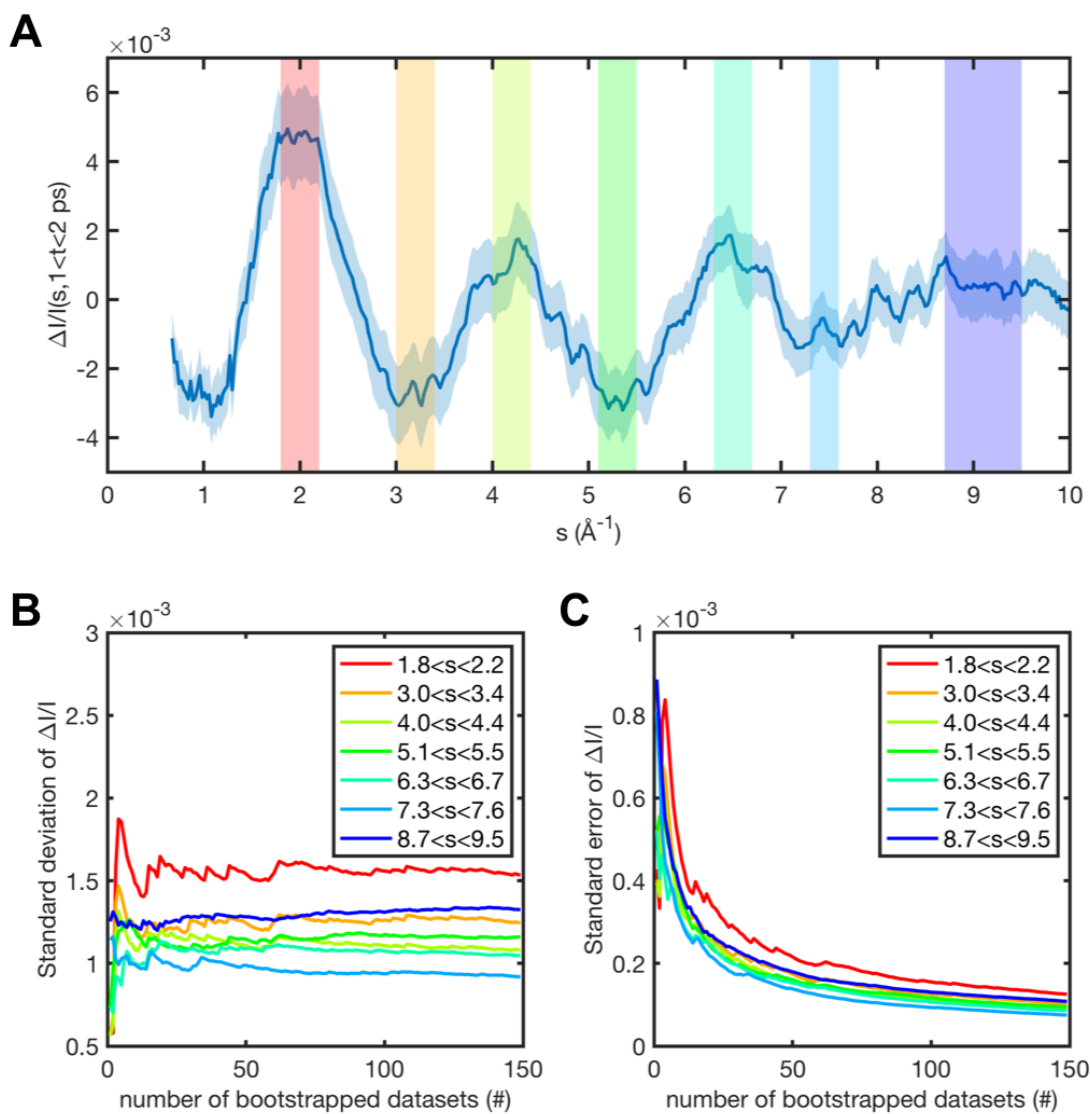

**Fig. S3: Uncertainty development through the dataset bootstrapping.** (**A**) shows the average experimental difference-diffraction signal for the time interval between 1 and 2 ps, with the shaded area representing one standard deviation across the 150 bootstrapped datasets. The shaded vertical bands represent the scattering features used to monitor the development of the uncertainty as a function of the number of bootstrapped datasets. (**B**) and (**C**) show, respectively, the development and saturation of the standard deviation and standard error of the difference-diffraction signal as a function of the number of bootstrapped datasets used in the analysis.

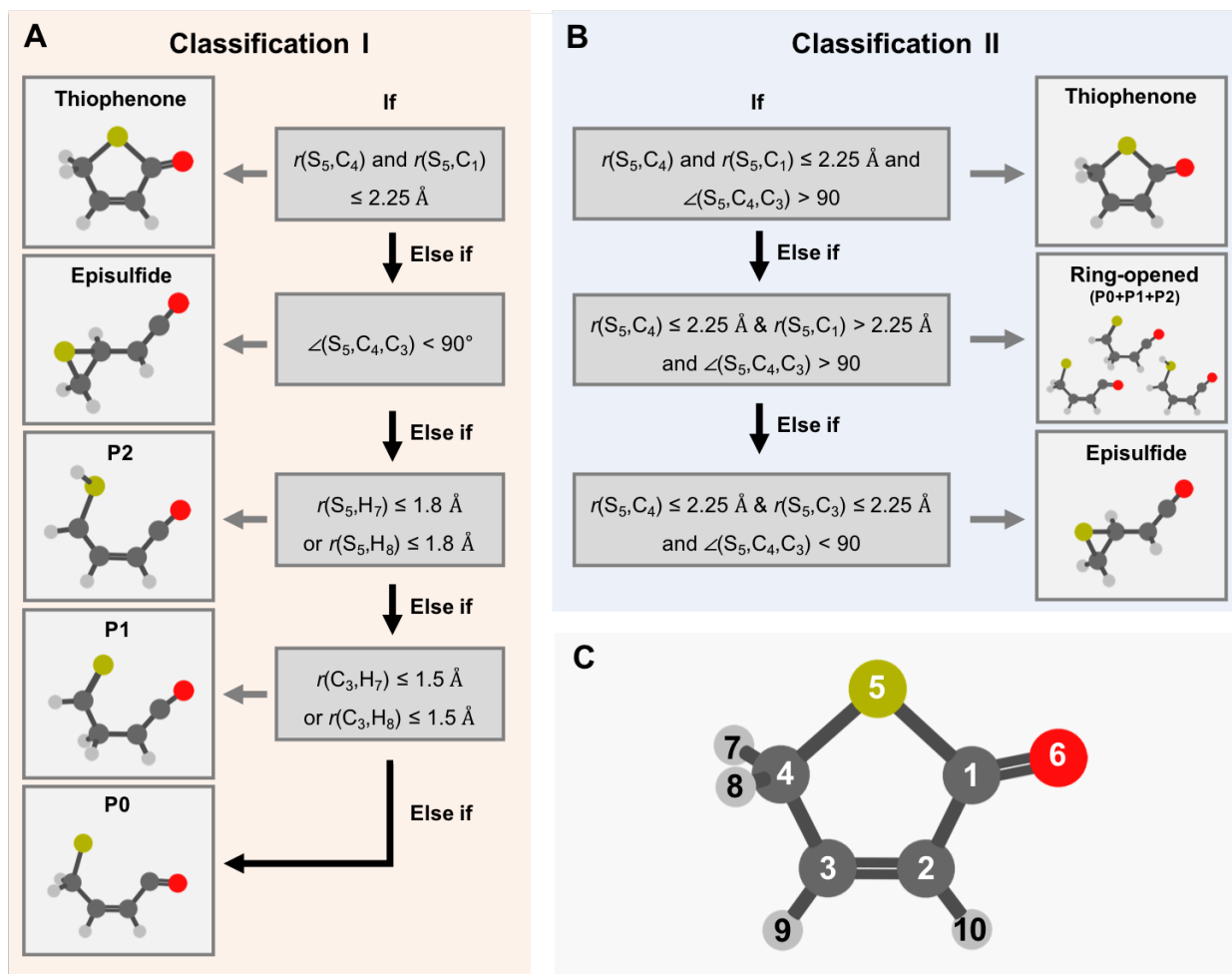

**Fig. S4: Decision trees for the classification of photoproducts.** (**A**): decision tree for classifying all photoproducts identified by our (NA+BO)MD simulations and reported in Ref. 11. (**B**): decision tree for the classification of photoproducts whose scattering signatures are distinct enough that their relative contributions can be unambiguously assigned using the experimental UED signal.

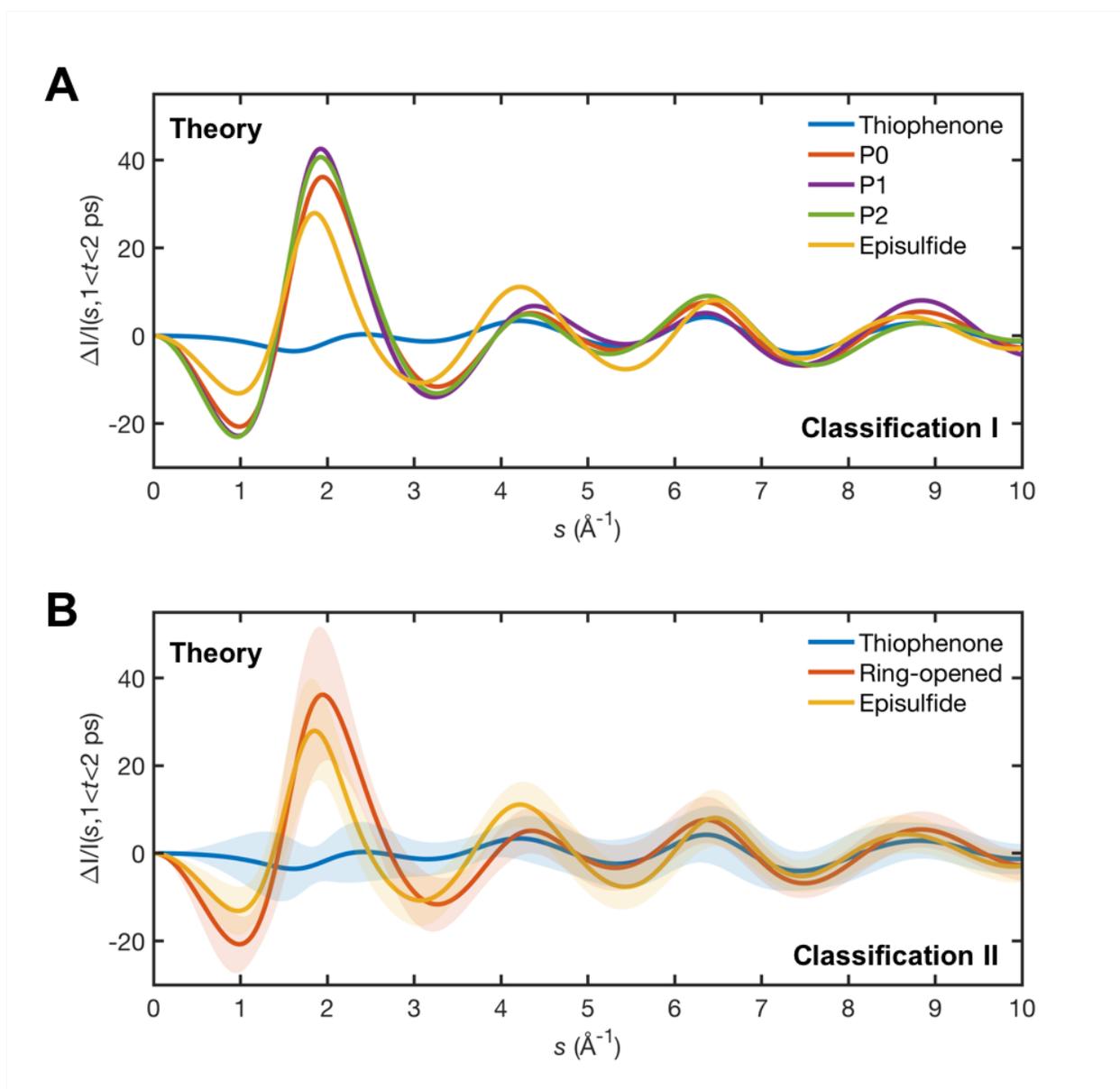

**Fig. S5: Photoproduct signatures according to classifications I and II.** (**A**): average scattering signatures of five photoproducts identified according to classification I (see Fig. S4). The scattering signatures of the ring-opened photoproducts P0, P1 and P2 are nearly indistinguishable (see Fig. S4 for the nomenclature of the photoproducts). Therefore, under classification II shown in (**B**), the relative scattering contributions from P0, P1 and P2 are grouped under a single category called *ring-opened*. The shaded regions in (**B**) represent one standard deviation across the ensemble of classified geometries. An equivalent representation is omitted from (**A**) for clarity purposes.

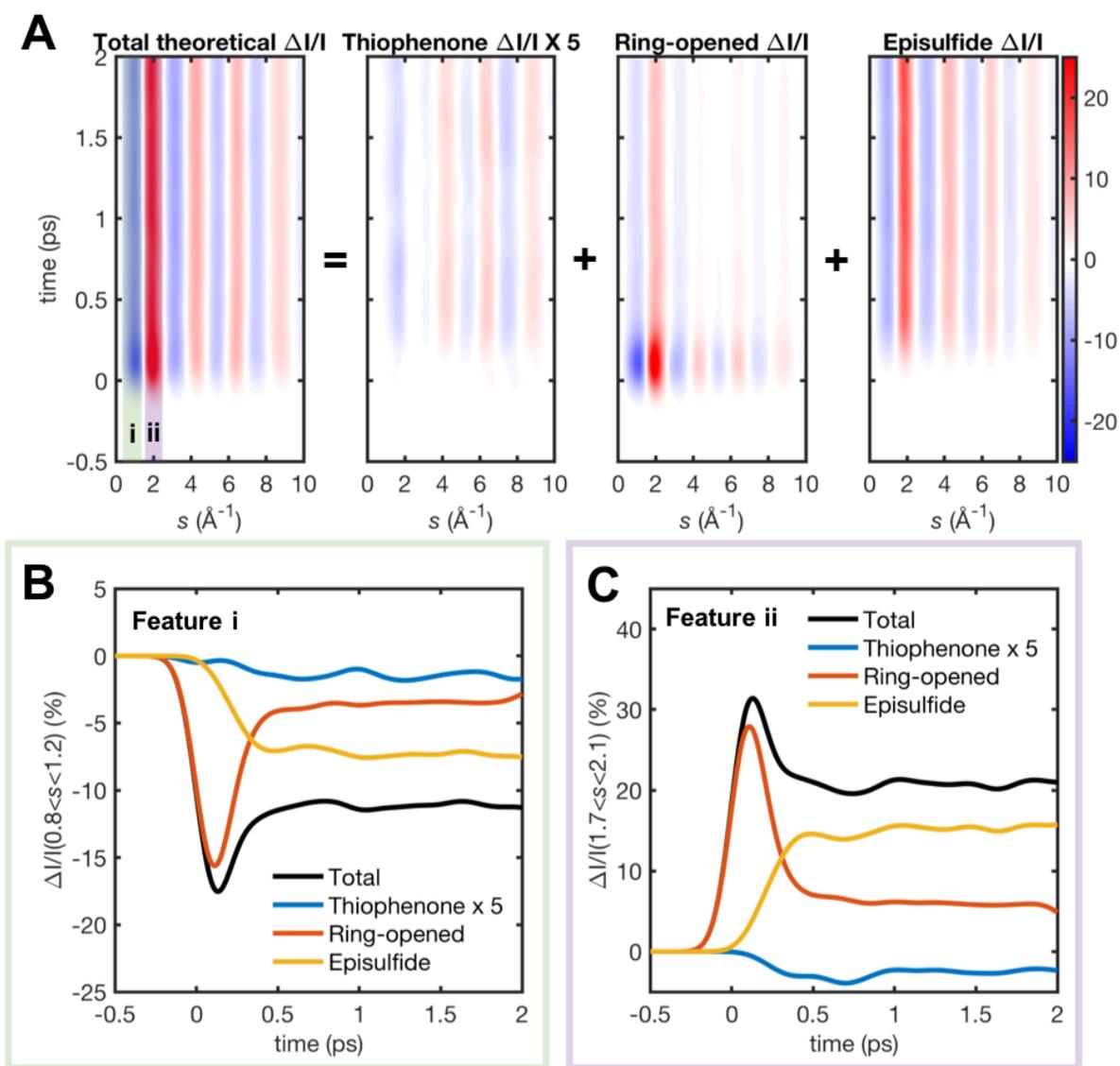

**Fig. S6: Photoproduct basis function selection.** (**A**): theoretical total $\Delta I/I(s,t)$ signal and relative contributions from each of the three photoproducts identified using classification II. These theoretical $\Delta I/I(s,t)$ maps have been convolved with a 230-fs FWHM Gaussian function which serves as an estimate for the instrument response function. (**B**) and (**C**): temporal evolution of the two strongest scattering features centered around: i) $+0.8 < s < +1.2$ and ii) $+1.7 < s < +2.1$ Å$^{-1}$, respectively. The temporal evolution of the total theoretical scattering signals and the relative contributions of each photoproduct reach a plateau after ~1 ps. The lack of large amplitude modulations on the temporal evolution of the photoproduct contributions to the total scattering signal at $t > 1$ ps enabled the use of time-independent photoproducts basis functions based on the average scattering signal in the range $+1 \leq t \leq +2$ ps.

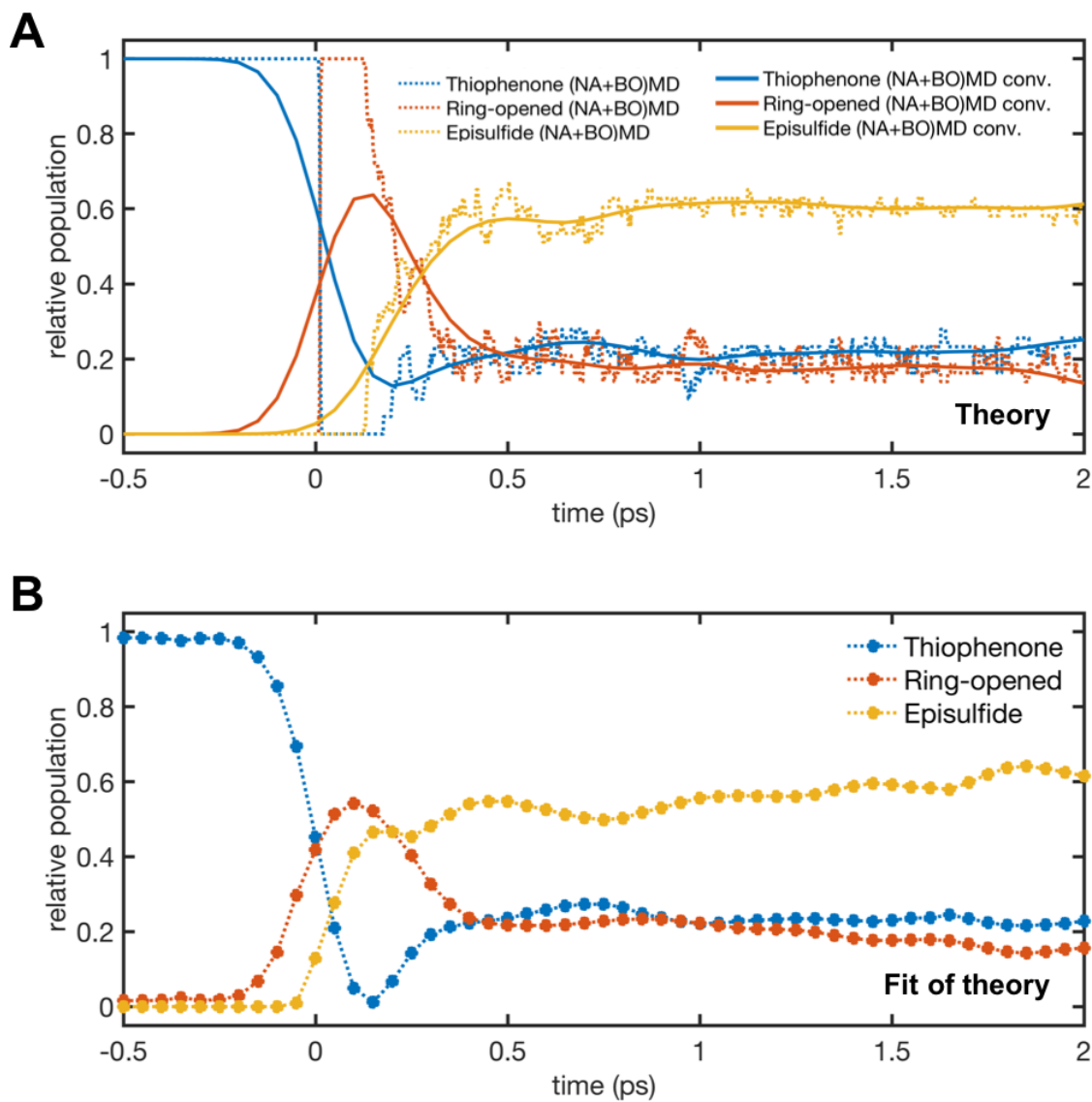

**Fig. S7: Benchmarking of the photoproduct relative population retrieval.** (**A**): temporal evolution of the photoproduct relative populations obtained from the (NA+BO)MD simulations directly (dashed lines). Each relative population in this case is defined as the fraction of trajectories, at a given time, entering a particular category identified according to classification II. The solid lines in (**A**) were obtained by convolving the time-dependent photoproduct relative populations with a 230-fs FWHM Gaussian function which approximates the experimental IRF. (**B**): temporal evolution of the photoproduct relative populations obtained by fitting the temporally convolved total theoretical signal, $\Delta I/I(s,t)$, with the time-independent basis functions shown in Fig. S5B.

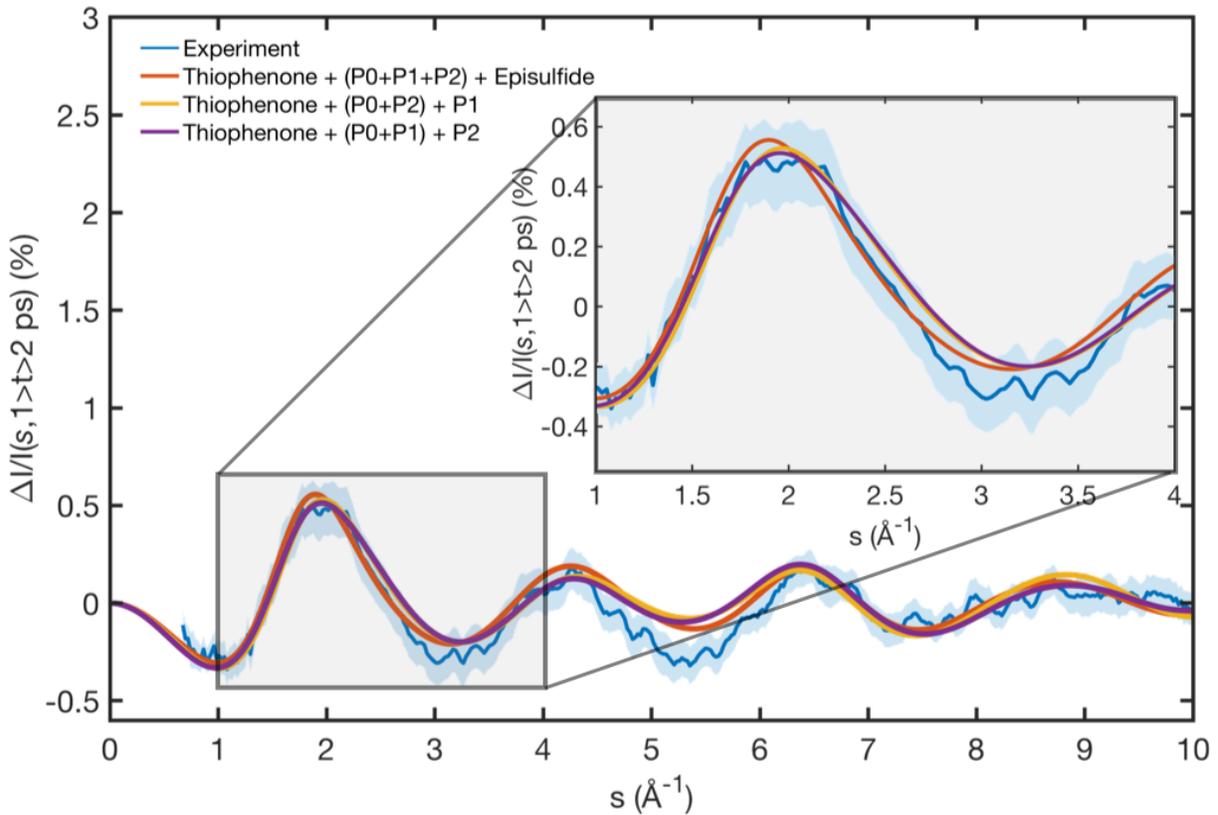

**Fig. S8: Benchmarking of basis function selection.** Fitting of the average experimental signal between 1 and 2 ps with three sets of basis functions including and excluding episulfide. The inset shows a zoom of the $1 < s < 4$ Å$^{-1}$ region. See Fig. S4 for the nomenclature of the photoproducts.

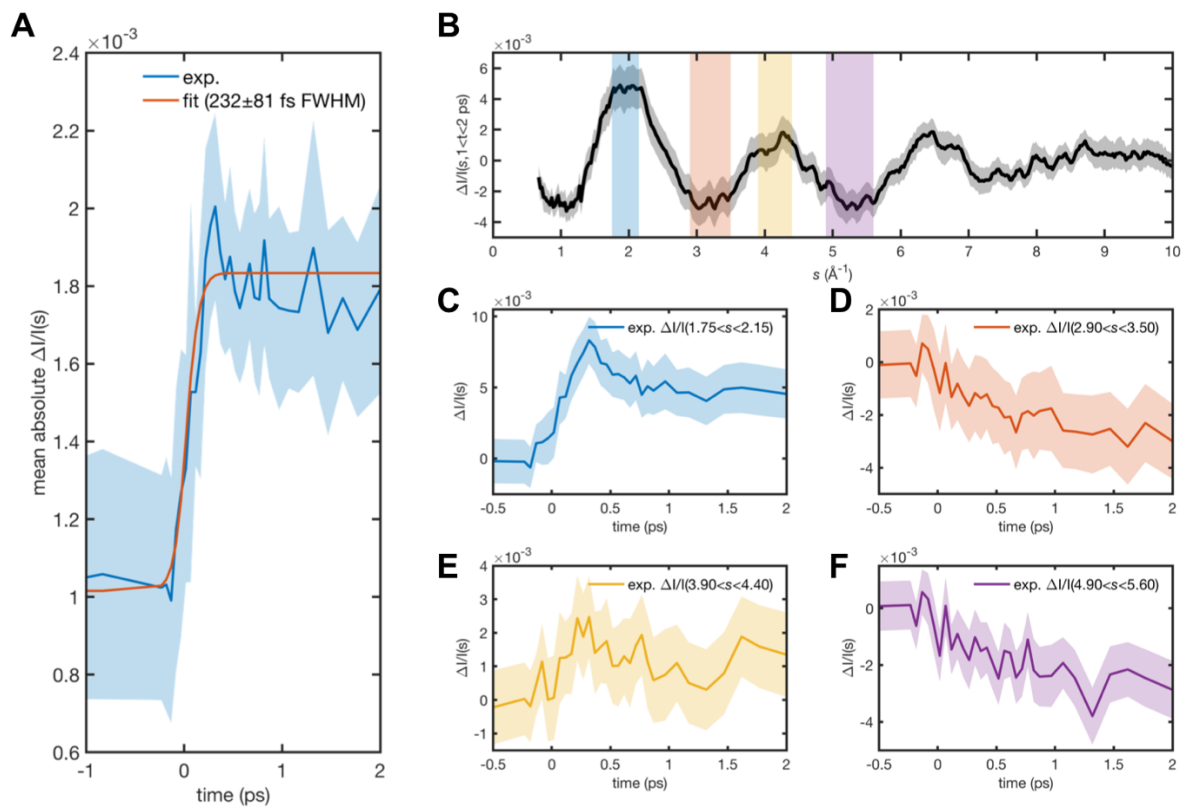

**Fig. S9: Experimental signal line-outs.** (**A**) shows the experimental average absolute difference-diffraction signal as function of time delay and corresponding Gaussian error fit which is used to estimate the instrument response function of the UED instrument. (**B**) shows the average experimental difference-diffraction signal for the time interval between 1 and 2 ps, with the shaded areas representing the four strongest features used to produce the line-outs in panels (**C**) to (**F**). The shaded areas in panels (**C**) to (**F**) represent one standard deviation across the 150 bootstrapped datasets.

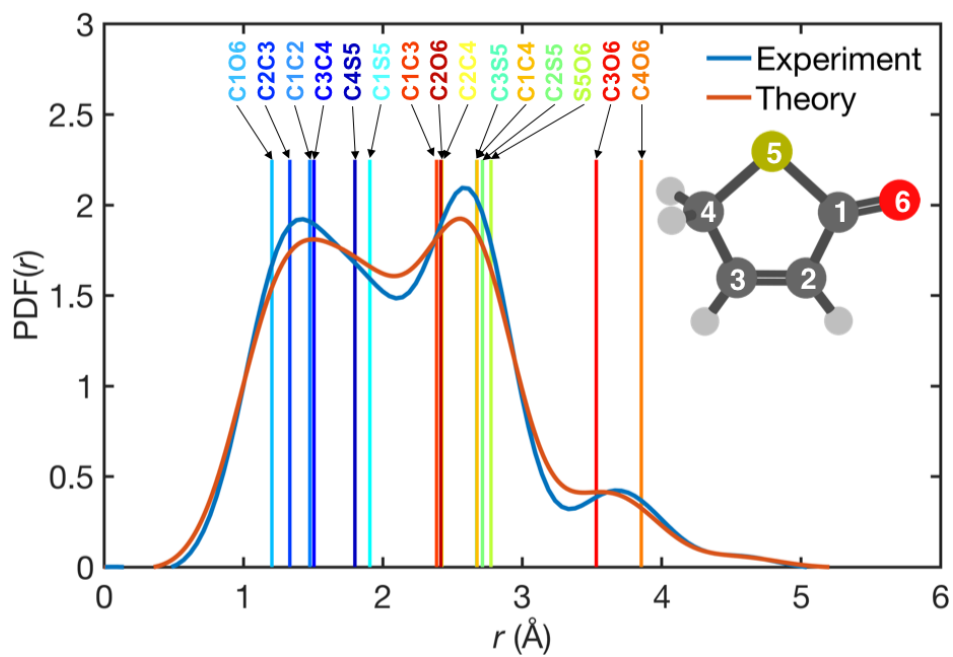

**Fig. S10: Assignment of the steady-state PDF for 2(5H)-thiophenone.** The vertical sticks represent average interatomic distances for heavy atoms obtained from the geometries sampled from the Wigner distribution.

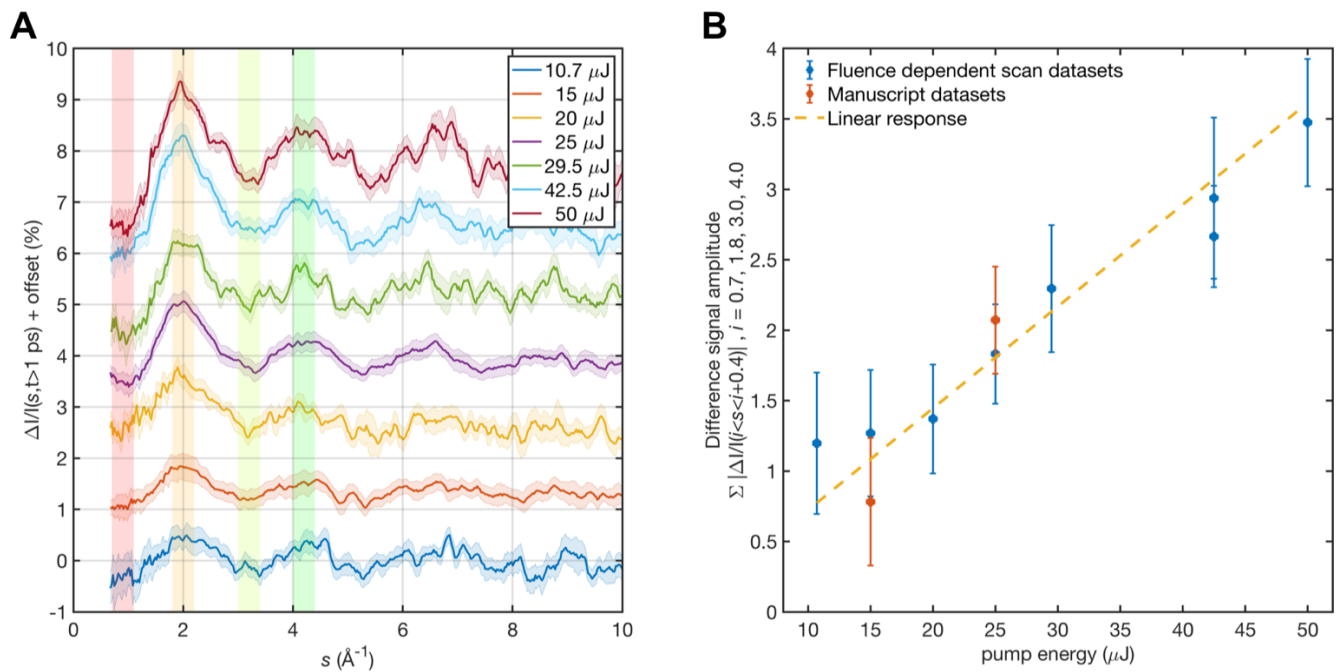

**Fig. S11: Fluence-dependence plot.** The shaded vertical bands in panel **A** represent the scattering features used to calculate the difference signal amplitude used in the fluence-dependence plot in panel **B**. Panel **B**: The red dot at 15 µJ corresponds to the dataset presented in the main text and the Supplementary Materials, and the red dot at 25 µJ is for the additional dataset analyzed in Fig. S12. We also note that the pump conditions in the present experiment were comparable to those in our prior photoelectron spectroscopy experiment on 2(5H)-thiophenone performed at FERMI (*10*), i.e., comparable UV pulse duration and focus size (200 µm at FERMI as opposed to 230 µm for MeV-UED) and slightly higher pulse energy (25 µJ at FERMI as compared to 15 µJ for the UED data presented in the main text). Since the FERMI experiment detected both photoelectrons and ions, it was possible to directly monitor the ionization of the target molecules by the pump pulse and to unambiguously determine that the degree of ionization under these conditions was negligible. The pump pulse energy of 25 µJ for the FERMI experiment was chosen because it was right below the onset of any noticeable ionization signal.

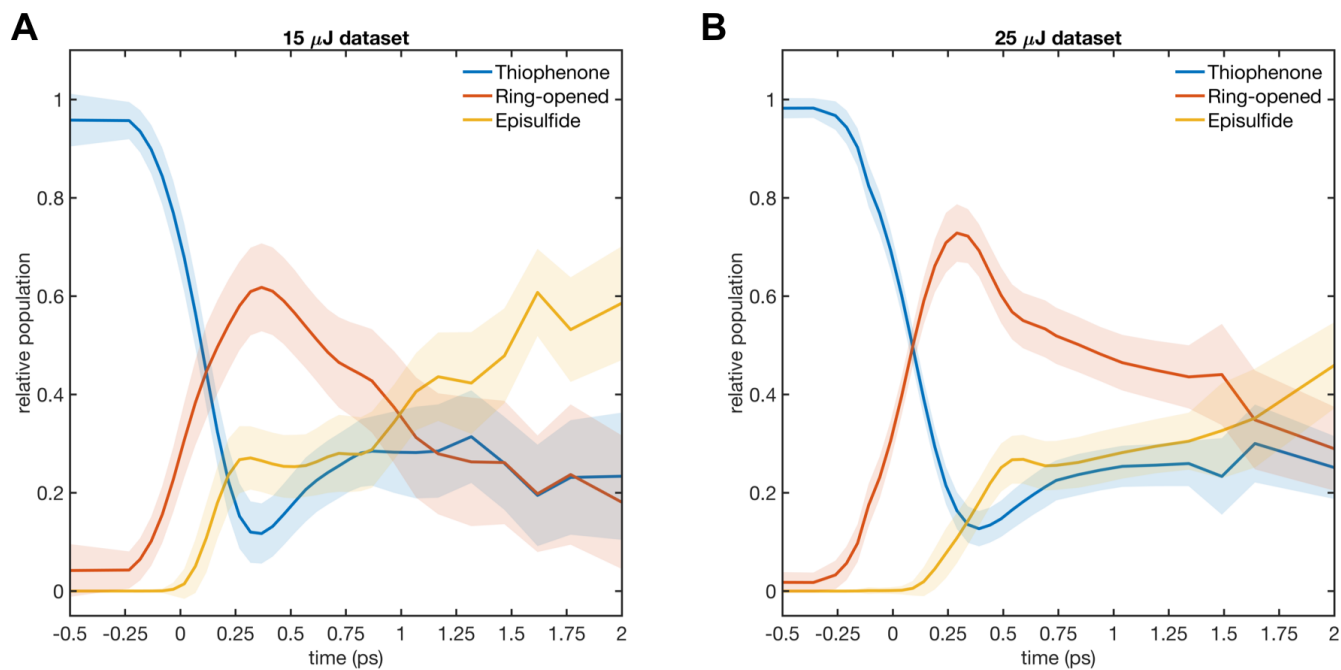

**Fig. S12: Time-resolved relative populations for the three families of photoproducts.** Time-dependent photoproduct populations extracted from datasets obtained with pump energies of 15 µJ (panel **A**, same as Fig. 5B in the main text) and 25 µJ (panel **B**). The two datasets shown here correspond to the two red dots in the panel **B** of Fig. S11.

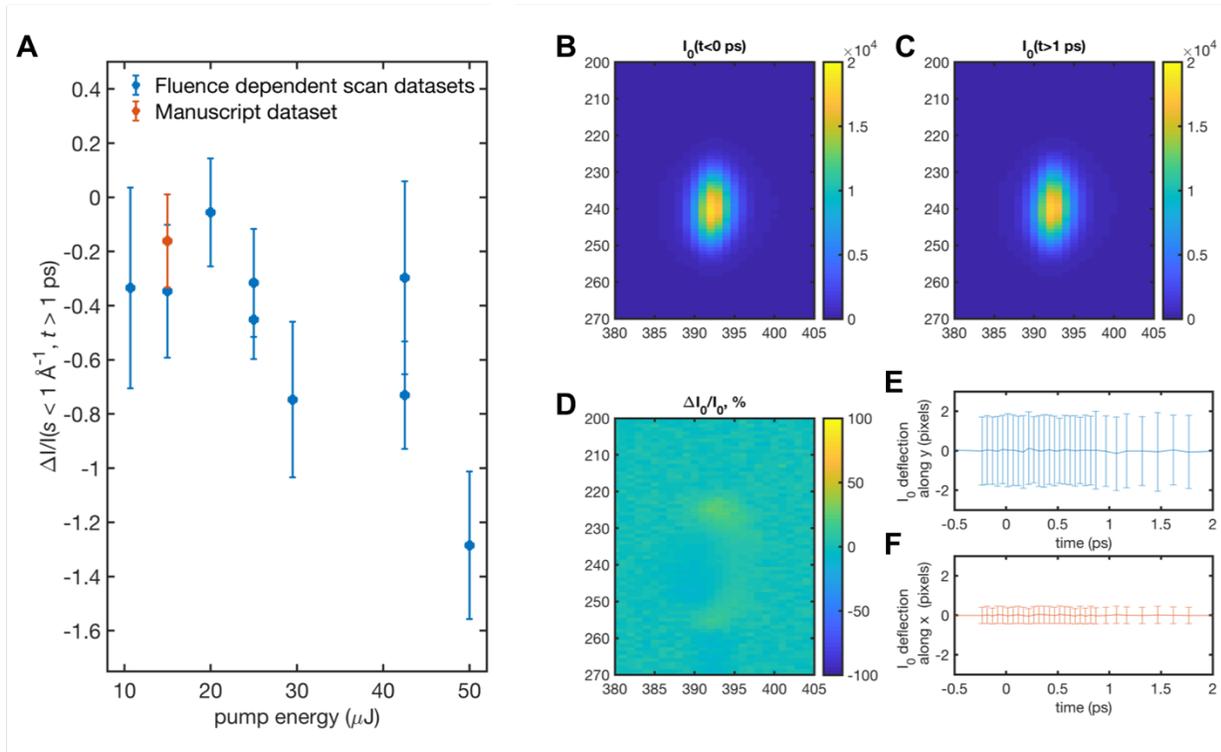

**Fig. S13: Assessment of the presence of photoionization induced contributions to the UED signal.** (**A**) shows a plot of the average difference signal at $s < 1$ Å$^{-1}$ and $t > 1$ picosecond as a function of pump energy. The red dot at 15 μJ corresponds to the dataset presented in the main text. (**B**) and (**C**) are spatially resolved false-color plots showing the average intensity of the undiffracted electron beam ($I_0$) before time-zero and after 1 picosecond, respectively. The intensity, shape and position of the undiffracted electron beam were measured on a 2-dimensional detector ($I_0$ detector) located immediately downstream of the main diffraction detector. The $I_0$ intensity is defined in detector counts by the false-color scale to the immediate right of panels B and C. The percentage difference signal between panels B and C is depicted in panel (**D**) and shows no appreciable lensing of the undiffracted electron beam. (**E**) and (**F**) show the position of the undiffracted electron beam centre-of-mass as a function of time along the x and y directions, respectively. We note that these data shown no time-dependent deflection of the undiffracted electron beam.

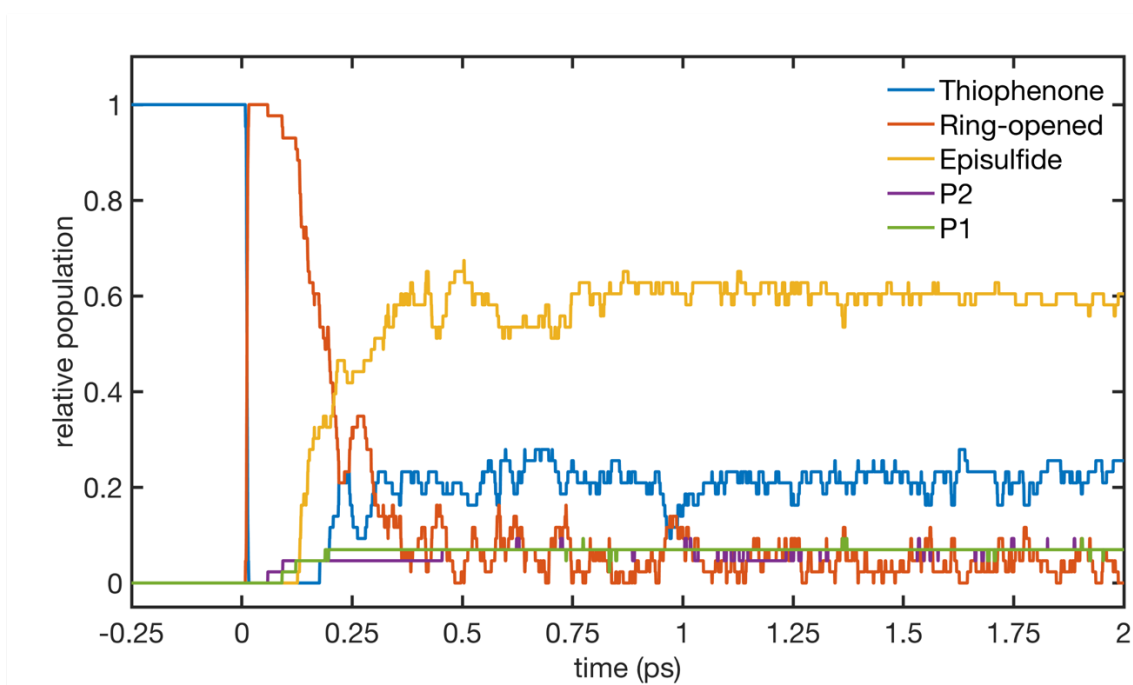

**Fig. S14: Time-resolved relative populations for the different photoproducts.** Temporal evolution of the photoproduct relative populations obtained from the (NA+BO)MD simulations, with the different ring-opened photoproducts – ring-opened, P1 and P2 – presented separately (see Fig. S4 for notation).

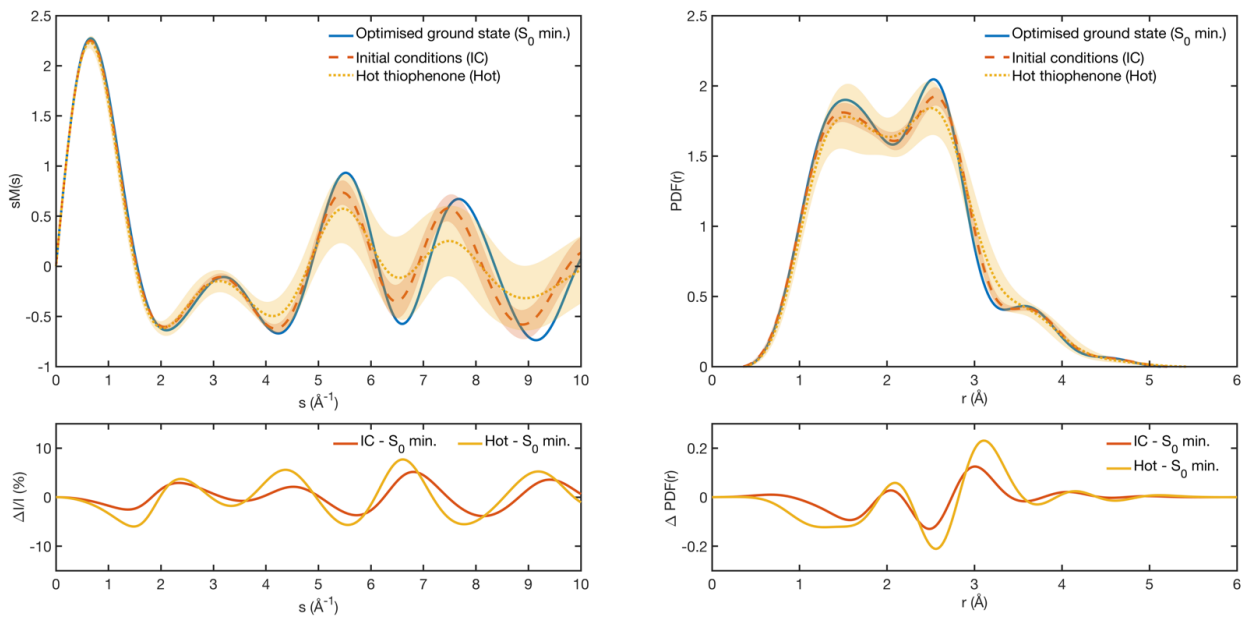

**Fig. S15: Simulated steady-state UED signals for 2(5H)-thiophenone from different representative structures.** Steady-state *sM*(*s*) (left) and PDF (right) for 2(5H)-thiophenone calculated from the ground-state optimized geometry of the molecule ($S_0$ min), the distribution of initial conditions sampled from a Wigner distribution mimicking the ground-state probability density at 0K (IC), and the distribution of 2(5H)-thiophenone geometries obtained from the (NA+BO)MD dynamics and representative of the ground-state athermal dynamics of (reformed) 2(5H)-thiophenone following the nonradiative decay (Hot). The lower panels show the Δ*I*/*I* and the ΔPDF obtained by subtracting the $S_0$ min signal.

**Table S1. Impact of basis function selection on the goodness of fit.**

Table showing the root mean square error for fits of the average experimental signal between 1 and 2 ps using different combinations of three photoproduct basis functions (see Fig. S4 for the nomenclature of the photoproducts).

| Basis Function | RMSE x $10^{-4}$ |
|---|---|
| 2(5H)-thiophenone + (P0+P1+P2) + Episulfide | 9.9±2.9 |
| 2(5H)-thiophenone + (P0+P2) + P1 | 11.0±2.7 |
| 2(5H)-thiophenone + (P0+P1) + P2 | 10.7±2.9 |
| Episulfide + (P0+P2) + P1 | 9.8±3.0 |
| Episulfide + (P0+P1) + P2 | 9.9±2.9 |
| P0 + P2 + P1 | 11.5±2.9 |

**Table S2. Impact of basis function selection and number on fit results.**

Table showing the relative populations, excitation percentage (exc.), and RMSE retrieved from fits of the average experimental signal between 1 and 2 ps using different combinations of photoproduct basis functions (see Fig. S4 for the nomenclature of the photoproducts).

| Basis functions | 2(5H)-thiophenone (%) | Ring-opened (%) | P2 (%) | P1 (%) | Episulfide (%) | exc. (%) | RMSE x $10^{-4}$ |
|---|---|---|---|---|---|---|---|
| 2(5H)-thiophenone + (P0+P1+P2) + Episulfide | 24±15 | 26±11 | - | - | 50±13 | 2.6±0.4 | 9.9±2.9 |
| 2(5H)-thiophenone + (P0+P2) + P1 + Episulfide | 19±15 | 29±12 | - | 0±1 | 51±12 | 2.5±0.4 | 9.6±2.9 |
| 2(5H)-thiophenone + (P0+P1) + P2 + Episulfide | 29±15 | 2±5 | 24±11 | - | 46±14 | 2.6±0.4 | 9.7±2.9 |